\documentclass[preprint,superscriptaddress,nofootinbib]{revtex4}
  
  \usepackage{amssymb,amsmath}
  \usepackage{graphicx}
  \begin{document}
  \title{$B^{\ast}$ ${\to}$ $\overline{D}D$ decays with perturbative QCD approach}
  \author{Junfeng Sun}
  \affiliation{Institute of Particle and Nuclear Physics,
              Henan Normal University, Xinxiang 453007, China}
  \author{Haiyan Li}
  \affiliation{Institute of Particle and Nuclear Physics,
              Henan Normal University, Xinxiang 453007, China}
  \author{Yueling Yang}
  \affiliation{Institute of Particle and Nuclear Physics,
              Henan Normal University, Xinxiang 453007, China}
  \author{Na Wang}
  \affiliation{Institute of Particle and Nuclear Physics,
              Henan Normal University, Xinxiang 453007, China}
  \author{Qin Chang}
  \affiliation{Institute of Particle and Nuclear Physics,
              Henan Normal University, Xinxiang 453007, China}
  \author{Gongru Lu}
  \affiliation{Institute of Particle and Nuclear Physics,
              Henan Normal University, Xinxiang 453007, China}
  \begin{abstract}
  The nonleptonic two-body $B^{\ast}$ ${\to}$ $\overline{D}D$
  weak decays are studied phenomenologically with the perturbative
  QCD factorization approach.
  It is found that the $B_{s}^{{\ast}0}$ ${\to}$ $D_{s}^{-}D_{s}^{+}$,
  $B_{d}^{{\ast}0}$ ${\to}$ $D_{d}^{-}D_{s}^{+}$, and
  $B_{u}^{{\ast}+}$ ${\to}$ $\overline{D}_{u}^{0}D_{s}^{+}$
  decays have branching ratios ${\gtrsim}$ $10^{-9}$, and
  might be promisingly measurable at the running LHC and
  forthcoming SuperKEKB experiments in the future.
  \end{abstract}
  \maketitle

  \section{Introduction}
  \label{sec01}
  The $B_{q}^{\ast}$ mesons, consisting of $\bar{b}q$ pair with
  $q$ $=$ $u$, $d$ and $s$, are spin-triplet ground vector states
  with definite spin-parity quantum numbers of $J^{P}$ $=$
  $1^{-}$ \cite{pdg}.
  Because the mass splittings $m_{B_{q}^{\ast}}$ $-$ $m_{B_{q}}$
  ${\lesssim}$ 50 MeV \cite{pdg} are much smaller than the mass of the
  lightest pion meson, the $B_{q}^{\ast}$ meson decays
  dominantly into the ground pseudoscalar $B_{q}$ meson through
  the electromagnetic interaction.
  Besides, the $B_{q}^{\ast}$ mesons can also decay via the
  bottom-changing transition induced by the weak interaction
  within the standard model (SM).
  Because of the strong phase-space suppression from their
  dominant magnetic dipole (M1) transition $B_{q}^{\ast}$
  ${\to}$ $B_{q}{\gamma}$,
  the lifetime of the $B_{q}^{\ast}$ meson is of the order of
  $10^{-17}$ second or less,
  which, in general, is too short to enable the $B_{q}^{\ast}$
  meson to experience the weak disintegration \cite{1509.05049}.
  The $B_{q}^{\ast}$ weak decays have not actually attracted
  much attention yet.
  Until now, there has been no experimental report and few
  theoretical works concentrating on the $B_{q}^{\ast}$ weak
  decay, subject to the relatively inadequate statistics on
  the $B_{q}^{\ast}$ mesons.
  Fortunately, the high luminosities and large production
  rates at LHC and the forthcoming SuperKEKB are
  promising, and the rapid accumulation of more and more
  $B_{q}^{\ast}$ data samples is expected to be possible.
  Some $B_{q}^{\ast}$ weak decay modes might be detected
  and investigated in the future, which undoubtedly makes the
  $B_{q}^{\ast}$ mesons another a vibrant arena for testing
  the Cabibbo-Kabayashi-Maskawa (CKM) picture for $CP$-violating
  phenomena, examining our comprehension of the underlying
  dynamical factorization mechanism, and so on.
  In addition, heavy quark symmetry relates hadronic transition
  matrix elements (HTME) of the $B_{q}^{\ast}$ and $B_{q}$ weak decays.
  The interplay between the $B_{q}^{\ast}$ and $B_{q}$ weak decays
  could prove useful information to overconstraint parameters in
  the SM, and might shed some fresh light on various anomalies in
  $B$ decays.

  The purely leptonic decays $B_{q}^{\ast}$ ${\to}$ ${\ell}^{+}{\ell}^{-}$
  induced by the flavor-changing neutral currents have been
  studied recently in the SM \cite{1509.05049,1601.03386}.
  The semileptonic and nonleptonic $B_{q}^{\ast}$ decays have
  been investigated also in the SM \cite{1605.01630,1605.01629,1605.01631,1606.09071},
  where the transition form factors are evaluated with the
  Wirbel-Stech-Bauer approach \cite{zpc29}, and the nonfactorizable
  corrections to HTME are considered \cite{1605.01629,1605.01631}
  based on the collinear-based and QCD-improved
  factorization (QCDF) approach \cite{qcdf1,qcdf2,qcdf3,plb.509.263,
  prd64.014036}.
  In this paper, we will study the nonleptonic $B_{q}^{\ast}$
  decay into the pseudoscalar charmed-meson pair $\overline{D}D$ with the
  perturbative QCD factorization (pQCD) approach \cite{pqcd1,pqcd2,pqcd3},
  just to provide a ready reference for the future experimental
  research.
  In addition, as is well known, the production ratio for
  the $B_{q}^{\ast}$ meson is comparable with that for the $B_{q}$
  meson (see Table \ref{tab:bb-fr}), the $B_{q}^{\ast}$ and
  $B_{q}$ mesons have nearly equal mass.
  Hence, the study of the $B_{q}^{\ast}$ ${\to}$ $\overline{D}D$
  decays will undoubtedly be helpful to the experimental background
  analysis on the $B_{q}$ ${\to}$ $\overline{D}D$ decays.

  This paper is organized as follows.
  The theoretical framework and amplitudes for $B_{q}^{\ast}$
  ${\to}$ $\overline{D}D$ decays with pQCD approach are given
  in section \ref{sec02}.
  Section \ref{sec03} is devoted to numerical results and discussion.
  The final section is a summary.

  \section{theoretical framework}
  \label{sec02}
  \subsection{The effective Hamiltonian}
  \label{sec0201}
  The effective Hamiltonian describing the $B_{q}^{\ast}$ ${\to}$
  $\overline{D}D$ weak decay is written as \cite{9512380}
   \begin{equation}
  {\cal H}_{\rm eff}\, =\, \frac{G_{F}}{\sqrt{2}}
   \sum\limits_{q=d,s} \Big\{
   \sum\limits_{p=u,c} V_{pb}^{\ast} V_{pq}
   \sum\limits_{i=1}^{2} C_{i}({\mu})\,Q_{i}({\mu})
   -V_{tb}^{\ast} V_{tq}\,
   \sum\limits_{k=3}^{10} C_{k}({\mu})\,Q_{k}({\mu}) \Big\}
   + {\rm h.c.}
   \label{hamilton},
   \end{equation}
  where the Fermi coupling constant $G_{F}$ ${\simeq}$
  $1.166{\times}10^{-5}\,{\rm GeV}^{-2}$ \cite{pdg};
  $V_{pb}^{\ast}V_{pq}$ and $V_{tb}^{\ast}V_{tq}$ are the CKM factors;
  The scale ${\mu}$ factorizes the physical contributions into two parts:
  the Wilson coefficients $C_{i}$ and the local four-fermion operators $Q_{i}$.
  The operators are defined as follows.
    \begin{eqnarray}
    Q_{1} &=&
   ( \bar{b}_{\alpha}\, p_{\alpha} )_{V-A}\,
   ( \bar{p}_{\beta}\, q_{\beta} )_{V-A},
     \qquad \qquad \quad \ \
    Q_{2}\ =\
   ( \bar{b}_{\alpha}\, p_{\beta} )_{V-A}\,
   ( \bar{p}_{\beta}\, q_{\alpha} )_{V-A}
    \label{q12}, \\
    Q_{3} &=& \sum\limits_{q^{\prime}}\,
   ( \bar{b}_{\alpha}\, q_{\alpha} )_{V-A}\,
   ( \bar{q}^{\prime}_{\beta}\, q^{\prime}_{\beta} )_{V-A},
     \qquad \qquad
   Q_{4} \ =\  \sum\limits_{q^{\prime}}\,
   ( \bar{b}_{\alpha}\, q_{\beta} )_{V-A}\,
   ( \bar{q}^{\prime}_{\beta}\, q^{\prime}_{\alpha} )_{V-A}
    \label{q34}, \\
    Q_{5} &=& \sum\limits_{q^{\prime}}\,
   ( \bar{b}_{\alpha}\, q_{\alpha} )_{V-A}\,
   ( \bar{q}^{\prime}_{\beta}\, q^{\prime}_{\beta} )_{V+A},
     \qquad \qquad
    Q_{6} \ =\ \sum\limits_{q^{\prime}}\,
   ( \bar{b}_{\alpha}\, q_{\beta} )_{V-A}\,
   ( \bar{q}^{\prime}_{\beta}\, q^{\prime}_{\alpha} )_{V+A}
    \label{q5}, \\
    Q_{7} &=& \sum\limits_{q^{\prime}}\, \frac{3}{2}\,Q_{q^{\prime}}\,
   ( \bar{b}_{\alpha}\, q_{\alpha} )_{V-A}\,
   ( \bar{q}^{\prime}_{\beta}\, q^{\prime}_{\beta} )_{V+A},
     \quad \
    Q_{8} \ =\ \sum\limits_{q^{\prime}}\, \frac{3}{2}\,Q_{q^{\prime}}\,
   ( \bar{b}_{\alpha}\, q_{\beta} )_{V-A}\,
   ( \bar{q}^{\prime}_{\beta}\, q^{\prime}_{\alpha} )_{V+A}
    \label{q78}, \\
    Q_{9} &=& \sum\limits_{q^{\prime}}\, \frac{3}{2}\,Q_{q^{\prime}}\,
   ( \bar{b}_{\alpha}\, q_{\alpha} )_{V-A}\,
   ( \bar{q}^{\prime}_{\beta}\, q^{\prime}_{\beta} )_{V-A},
     \quad
   Q_{10} \ =\ \sum\limits_{q^{\prime}}\, \frac{3}{2}\,Q_{q^{\prime}}\,
   ( \bar{b}_{\alpha}\, q_{\beta} )_{V-A}\,
   ( \bar{q}^{\prime}_{\beta}\, q^{\prime}_{\alpha} )_{V-A}
    \label{q90},
    \end{eqnarray}
  where $Q_{1,2}$ are tree operators arising from the $W$-boson
  exchange; $Q_{3,{\cdots},6}$ and $Q_{7,{\cdots},10}$ are called
  the QCD and electroweak penguin operators, respectively;
  $(\bar{q}_{1}\,q_{2})_{V{\pm}A}$ ${\equiv}$
  $\bar{q}_{1}\,{\gamma}_{\mu}(1{\pm}{\gamma}_{5})\,q_{2}$;
  ${\alpha}$ and ${\beta}$ are color indices;
  $q^{\prime}$ denotes all the active quarks at the scale of
  ${\cal O}(m_{b})$, i.e., $q^{\prime}$ $=$ $u$, $d$, $c$, $s$, $b$;
  and $Q_{q^{\prime}}$ is the electric charge of quark $q^{\prime}$
  in the unit of ${\vert}e{\vert}$.

  The Wilson coefficients $C_{i}(\mu)$, which summarize the
  physical contributions above the scale of ${\mu}$, have
  been properly calculated at the next-to-leading order with
  the renormalization group equation assisted perturbation
  theory \cite{9512380}.
  Due to the presence of long-distance QCD effects and the
  entanglement of nonperturbative and perturbative shares,
  the main obstacle to evaluate the $B_{q}^{(\ast)}$ weak decays
  is the treatment of physical contributions below the
  scale of ${\mu}$ which are included in the HTME of local operators.

  \subsection{Hadronic matrix elements}
  \label{sec0202}
  Some phenomenological models have recently been developed to improve
  the sketchy treatment with naive factorization scheme \cite{npb133,zpc34}.
  These models are generally based on the Lepage-Brodsky approach
  \cite{prd22} and some power counting rules in parameters of
  ${\alpha}_{s}$ and ${\Lambda}_{\rm QCD}/m_{Q}$ (where ${\alpha}_{s}$
  is the strong coupling, ${\Lambda}_{\rm QCD}$ is the QCD characteristic
  scale, and $m_{Q}$ is the mass of a heavy quark), and express the
  HTME as a convolution integral of universal wave functions
  and hard scattering subamplitudes, such as the QCDF approach
  \cite{qcdf1,qcdf2,qcdf3}, pQCD approach \cite{pqcd1,pqcd2,pqcd3},
  the soft and collinear effective theory \cite{scet1,scet2,scet3,scet4},
  and so on, which have been extensively employed in the interpretation
  of the $B$ weak decays.
  To wipe out the endpoint singularities appearing in the collinear
  approximation \cite{qcdf1,qcdf2,qcdf3},
  it is suggested by the pQCD approach \cite{pqcd1,pqcd2,pqcd3}
  that the transverse momentum $k_{T}$ of valence quarks should be
  retaken, and a Sudakov factor should be introduced for each wave
  function to further suppress the soft contributions and make the
  hard scattering more perturbative.
  Finally, a decay amplitude is written as a multidimensional
  integral of many parts \cite{pqcd2,pqcd3}, including the Wilson
  coefficients $C_{i}$, the heavy quark decay subamplitudes
  ${\cal H}$, and the universal wave functions ${\Phi}$,
  \begin{equation}
 {\cal A}\, {\sim}\,
  \sum_{i} {\int} {\prod_j}dk_{j}\,
  C_{i}(t)\,{\cal H}_{i}(t,k_{j})\,{\Phi}_{j}(k_{j})\,e^{-S_{j}}
  \label{hadronic},
  \end{equation}
  where $t$ is a typical scale; $k_{j}$ is the momentum of a valence
  quark; $e^{-S_{j}}$ is a Sudakov factor.

  \subsection{Kinematic variables}
  \label{sec0203}
\begin{figure}[h]
  \includegraphics[width=0.9\textwidth,bb=85 400 525 720]{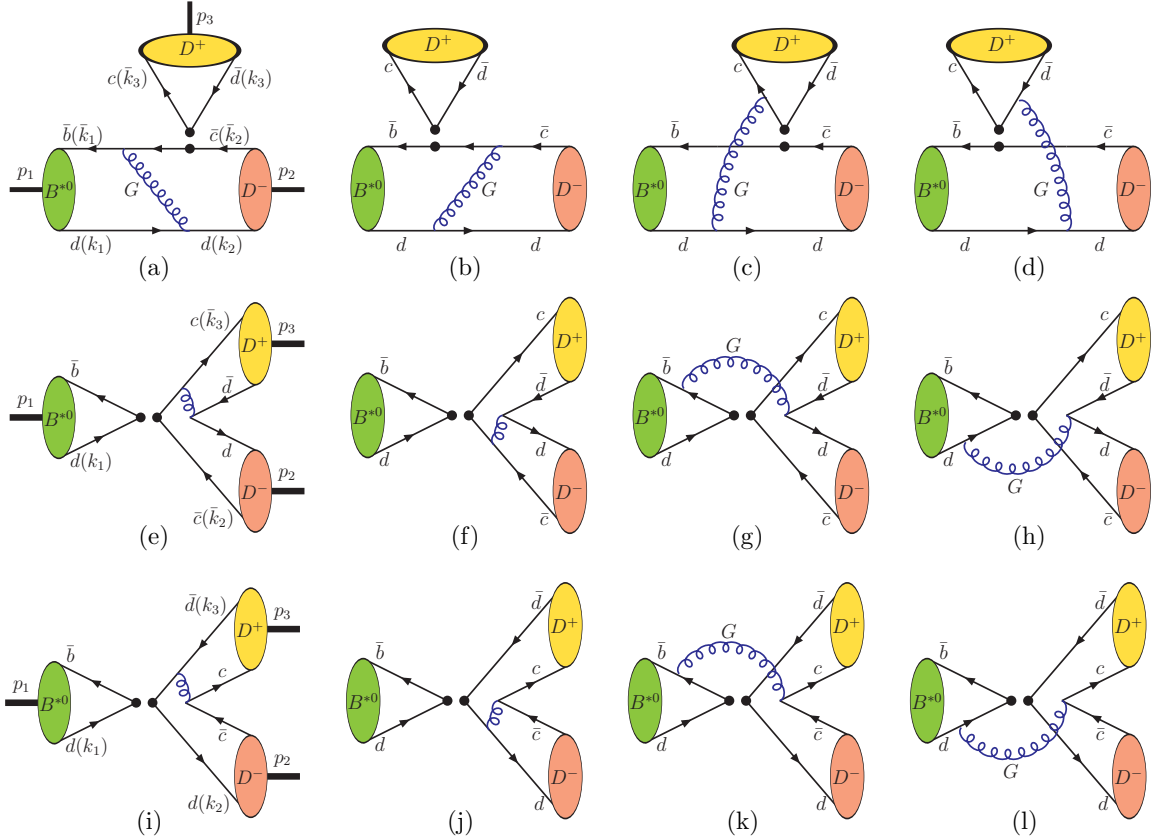}
  \caption{Feynman diagrams for $B^{{\ast}0}$ ${\to}$ $D^{+}D^{-}$
   decay, where (a,b) are the factorizable emission topologies,
   (c,d) are the nonfactorizable emission topologies,
   (e,f,i,j) are the factorizable annihilation topologies,
   and (g,h,k,l) are the nonfactorizable annihilation topologies.}
  \label{feynman}
  \end{figure}

  The light-cone variables in the rest frame of the $B^{\ast}$ meson
  are defined as follows.
  \begin{equation}
  p_{B^{\ast}}\, =\, p_{1}\, =\, \frac{m_{1}}{\sqrt{2}}(1,1,0)
  \label{kine-p1},
  \end{equation}
  \begin{equation}
  p_{\overline{D}}\, =\, p_{2}\, =\, (p_{2}^{+},p_{2}^{-},0)
  \label{kine-p2},
  \end{equation}
  \begin{equation}
  p_{D}\, =\, p_{3}\, =\, (p_{3}^{-},p_{3}^{+},0)
  \label{kine-p3},
  \end{equation}
  \begin{equation}
  k_{i}\, =\, x_{i}\,p_{i}+(0,0,\vec{k}_{i{T}})
  \label{kine-ki},
  \end{equation}
  \begin{equation}
  p_{i}^{\pm}\, =\, (E_{i}\,{\pm}\,p)/\sqrt{2}
  \label{kine-pipm},
  \end{equation}
  \begin{equation}
  {\epsilon}_{B^{\ast}}^{{\parallel}}\, =\, \frac{1}{\sqrt{2}}(-1,1,0)
  \label{kine-e1},
  \end{equation}
  \begin{equation}
  s\, =\, 2\,p_{2}{\cdot}p_{3}
  \label{kine-s},
  \end{equation}
  \begin{equation}
  t\, =\, 2\,p_{1}{\cdot}p_{2}\, =\ 2\,m_{1}\,E_{2}
  \label{kine-t},
  \end{equation}
  \begin{equation}
  u\, =\, 2\,p_{1}{\cdot}p_{3}\, =\ 2\,m_{1}\,E_{3}
  \label{kine-u},
  \end{equation}
  \begin{equation}
  s\,t +s\,u-t\,u -4\,m_{1}^{2}\,p^{2}\ =\ 0
  \label{kine-pcm},
  \end{equation}
  where the subscripts $i$ $=$ $1$, $2$, $3$ of variables (energy $E_{i}$,
  momentum $p_{i}$ and mass $m_{i}$) correspond to $B^{\ast}$, $\overline{D}$
  and $D$ mesons, respectively; $k_{i}$ is the momentum of the valence quark
  with the longitudinal momentum fraction $x_{i}$ and the transverse momentum
  $\vec{k}_{iT}$; ${\epsilon}_{B^{\ast}}^{{\parallel}}$ is the longitudinal
  polarization vector; $p$ is the momentum of the final
  states; $s$, $t$ and $u$ are the Lorentz invariant parameters.
  The notation is displayed in Fig.\ref{feynman}.

  \subsection{Wave functions}
  \label{sec0204}
  Wave functions are the basic input parameters with the pQCD approach.
  Although wave functions contain soft and nonperturbative contributions,
  they are universal, i.e., process independent.
  Wave functions and/or distribution amplitudes (DAs) determined
  by nonperturbative methods or extracted from data, can be employed
  here to make predictions.

  Following the notations in Refs. \cite{npb529,prd65,plb752sun,ijmpa31yang},
  HTME of the diquark operators is defined as
  \begin{equation}
 {\langle}0{\vert}\bar{b}_{i}(0)q_{j}(z){\vert}B^{\ast}(p,{\epsilon}^{\parallel}){\rangle}\,
 =\, \frac{f_{B^{\ast}}}{4} {\int}d^{4}k\,e^{-ik{\cdot}z}
  \Big\{ \!\!\not{\!\epsilon}^{\parallel}\, \Big[
  m_{B^{\ast}}\, {\Phi}_{B^{\ast}}^{v}(k) -
  \!\!\not{p}\, {\Phi}_{B^{\ast}}^{t}(k) \Big] \Big\}_{ji}
  \label{diquark-bs01},
  \end{equation}
  \begin{equation}
 {\langle}\overline{D}(p){\vert}c_{i}(0)\bar{q}_{j}(z){\vert}0{\rangle}\,
 =\, \frac{i\,f_{D}}{4}{\int}d^{4}k\,e^{+ik{\cdot}z}\,
  \Big\{ {\gamma}_{5}\, \Big[ \!\!\not{p}\,{\Phi}_{\bar{D}}^{a}(k)
  +m_{\bar{D}}\,{\Phi}_{\bar{D}}^{p}(k) \Big] \Big\}_{ji}
  \label{diquark-ds01},
  \end{equation}
  where $f_{B^{\ast}}$ and $f_{D}$ are decay constants; the
  wave functions ${\Phi}_{B^{\ast}}^{v}$ and ${\Phi}_{\bar{D}}^{a}$
  are twist-2; and ${\Phi}_{B^{\ast}}^{t}$ and ${\Phi}_{\bar{D}}^{p}$
  are twist-3.
  Due to the kinematic relation ${\epsilon}_{B^{\ast}}^{\perp}{\cdot}p_{i}$
  $=$ $0$, the transversely polarized $B^{\ast}$ meson contributes
  nothing to the amplitudes for the $B^{\ast}$ ${\to}$ $\overline{D}D$ decays.
  The expressions for DAs of the $B^{\ast}$ and $D$ mesons are
  \cite{plb752sun,ijmpa31yang}
   \begin{equation}
  {\phi}_{B^{\ast}}^{v}(x) = A\, x\,\bar{x}\,
  {\exp}\Big\{ -\frac{\bar{x}\,m_{q}^{2}+x\,m_{b}^{2}}
                     {8\,{\omega}_{1}^{2}\,x\,\bar{x}} \Big\}
   \label{da-bqv},
   \end{equation}
   \begin{equation}
  {\phi}_{B^{\ast}}^{t}(x) = B\, (\bar{x}-x)^{2}\,
  {\exp}\Big\{ -\frac{\bar{x}\,m_{q}^{2}+x\,m_{b}^{2}}
                     {8\,{\omega}_{1}^{2}\,x\,\bar{x}} \Big\}
   \label{da-bqt},
   \end{equation}
   \begin{equation}
  {\phi}_{\bar{D}}^{a}(x) = C\, x\,\bar{x}\,
  {\exp}\Big\{ -\frac{\bar{x}\,m_{q}^{2}+x\,m_{c}^{2}}
                     {8\,{\omega}_{2}^{2}\,x\,\bar{x}} \Big\}
   \label{da-cqa},
   \end{equation}
   \begin{equation}
  {\phi}_{\bar{D}}^{p}(x) = D\,
  {\exp}\Big\{ -\frac{\bar{x}\,m_{q}^{2}+x\,m_{c}^{2}}
                     {8\,{\omega}_{2}^{2}\,x\,\bar{x}} \Big\}
   \label{da-cqp},
   \end{equation}
   where $x$ and $\bar{x}$ (${\equiv}$ $1$ $-$ $x$) are the
   parton momentum fractions;
   ${\omega}_{i}$ determines the average transverse momentum
   of partons and ${\omega}_{i}$ ${\simeq}$ $m_{i}\,{\alpha}_{s}(m_{i})$;
   parameters $A$, $B$, $C$, $D$ are the normalization coefficients
   satisfying the conditions
   \begin{equation}
  {\int}_{0}^{1}dx\,{\phi}_{B^{\ast}}^{v,t}(x)=1
   \label{wave-nb},
   \end{equation}
   \begin{equation}
  {\int}_{0}^{1}dx\,{\phi}_{\bar{D}}^{a,p}(x) =1
   \label{wave-nd}.
   \end{equation}

  In fact, there are many phenomenological models of DAs for the
  charmed meson, for example, some of them have been listed by
  Eq.(30) in  Ref.\cite{prd78.014018}.
  One of the favorable models from the experimental data within the pQCD
  framework has the expression \cite{prd78.014018}
   \begin{equation}
  {\phi}_{D}(x) = 6\,x\bar{x}\,\big\{1+C_{D}(1-2x) \big\}
   \label{wave-d-xb},
   \end{equation}
  where parameter $C_{D}$ $=$ $0.5$ for the $D_{u,d}$ meson,
  and $C_{D}$ $=$ $0.4$ for the $D_{s}$ meson.
  In the actual calculation \cite{prd78.014018,prd81.034006,jpg31.273,jpg37},
  there is no distinction between twist-2 and twist-3 DAs.

\begin{figure}[h]
  \includegraphics[width=0.8\textwidth,bb=75 570 540 720]{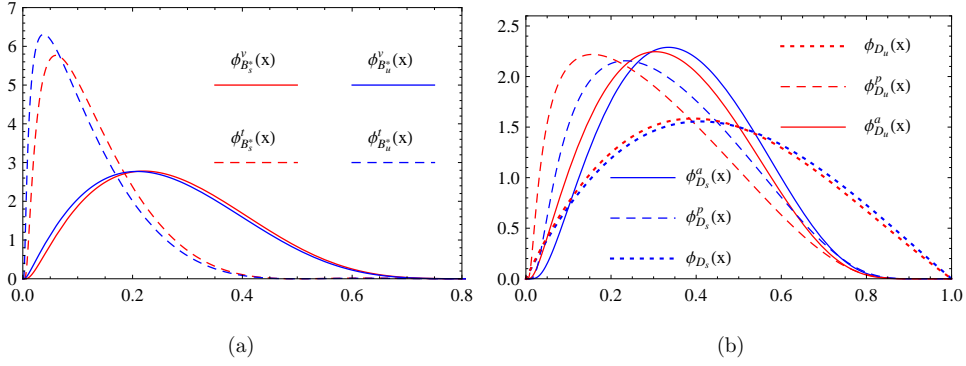}
  \caption{The distributions of DAs ${\phi}_{B^{\ast}}^{v,t}(x)$
  [Eqs.(\ref{da-bqv},\ref{da-bqt})], ${\phi}_{D}^{a,p}(x)$
  [Eqs.(\ref{da-cqa},\ref{da-cqp})], and ${\phi}_{D}(x)$
  [Eq.(\ref{wave-d-xb})].}
  \label{fig:da}
  \end{figure}

  The shape lines of the normalized DAs ${\phi}_{B^{\ast}}^{v,t}(x)$
  and ${\phi}_{D}^{a,p}(x)$ are illustrated in Fig.\ref{fig:da}.
  It is clearly seen from Fig.\ref{fig:da} that
  (1) the shape lines of DAs in Eqs.(\ref{da-bqv})-(\ref{da-cqp}) have a
  broad peak in the small $x$ regions, which is generally consistent
  with an ansatz in which a light quark carries fewer parton momentum
  fractions than a heavy quark in a heavy-light system.
  (2) Due to the suppression from exponential functions, the DAs of
  Eqs.(\ref{da-bqv})-(\ref{da-cqp}) converge quickly to zero
  at endpoint $x$, $\bar{x}$ ${\to}$ $0$, which supplies the
  soft contributions with an effective cutoff.
  (3) The flavor symmetry breaking effects,
  and especially the distinction between different the twist DAs, are
  highlighted, compared with the nearly symmetric distribution
  Eq.(\ref{wave-d-xb}).


  \subsection{Decay amplitudes}
  \label{sec0205}
  The Feynman diagrams for the $B^{{\ast}0}$ ${\to}$ $D^{+}D^{-}$ decay
  with the pQCD approach are shown in Fig.\ref{feynman}, including the
  factorizable emission topologies (a,b) where one gluon links with
  the initial and the recoiled states, the
  nonfactorizable emission topologies (c,d) where one gluon is exchanged
  between the spectator quark and the emitted states, the
  factorizable annihilation topologies (e,f,i,j) where one gluon
  is conjoined with the final states, and the nonfactorizable annihilation
  topologies (g,h,k,l) where one gluon is transmitted between the
  initial and the final states.

  Generally, the amplitudes for the factorizable emission
  topologies in Fig.\ref{feynman}(a,b) can be written as the $D$
  meson decay constant and the space-like $B^{\ast}$ ${\to}$ $D$
  transition form factor, and the amplitudes for the factorizable
  annihilation topologies in Fig.\ref{feynman}(e,f,i,j) can be
  written as the $B^{\ast}$ meson decay constant and the time-like
  transition form factor between two charmed mesons.
  The amplitudes for the nonfactorizable topologies have more
  complicated structures, and can be written as the convolution
  integral of all participating meson wave functions.
  Compared with the contributions of the emission topologies
  in Fig.\ref{feynman}(a-d), the contributions of the annihilation
  topologies in Fig.\ref{feynman}(e-l) are assumed to be power
  suppressed, as stated in Ref.\cite{qcdf2}.
  In addition, different topologies have different scales.
  The gluons of the emission topologies in Fig.\ref{feynman}(a-d)
  are time-like, while the gluons of the annihilation topologies
  in Fig.\ref{feynman}(e-l) are space-like.
  The gluon virtuality of creating a pair of heavy charm quarks
  from the vacuum for the annihilation topologies in Fig.\ref{feynman}(i-l),
  $k_{g}^{2}$ ${\ge}$ $(2m_{c})^{2}$, should be much larger than
  that of producing a pair of light quarks for the annihilation
  topologies in Fig.\ref{feynman}(e-h).
  Thus, it is not hard to figure out that the contributions of the
  annihilation topologies in Fig.\ref{feynman}(i-l) might be very
  small relative to the others, because of the nature of the
  asymptotic freedom of the QCD at the unltrahigh energy.

  After a straightforward calculation, the amplitudes for the $B_{q}^{\ast}$
  ${\to}$ $\overline{D}D$ decays are expressed as below.
   \begin{eqnarray} & &
  {\cal A}(B^{{\ast}+}{\to}\overline{D}^{0}D_{q}^{+})
   \nonumber \\ &=& {\cal F}\, \Big\{
   V_{cb}^{\ast}\,V_{cq}\,
   \Big[ a_{1}\, {\cal A}_{a+b}^{LL}+C_{2}\, {\cal A}_{c+d}^{LL} \Big]
  +V_{ub}^{\ast}\,V_{uq}\,
   \Big[ a_{1}\, {\cal A}_{i+j}^{LL}+C_{2}\, {\cal A}_{k+l}^{LL} \Big]
   \nonumber \\ & & \quad
  -V_{tb}^{\ast}\,V_{tq}\,
   \Big[ (a_{4}+a_{10})\, {\cal A}_{a+b+i+j}^{LL}
        +(a_{6}+a_{8})\, {\cal A}_{a+b}^{SP}
   \nonumber \\ & & \qquad \qquad
      +\ (C_{3}+C_{9})\, {\cal A}_{c+d+k+l}^{LL}
        +(C_{5}+C_{7})\, {\cal A}_{c+d+k+l}^{SP} \Big] \Big\}
   \label{eq:amp-bu},
   \end{eqnarray}
   \begin{eqnarray} & &
  {\cal A}(B_{q}^{{\ast}0}{\to}\overline{D}^{0}D^{0})
   \nonumber \\ &=& {\cal F}\, \Big\{
   V_{cb}^{\ast}\,V_{cq}\,
   \Big[ a_{2}\, {\cal A}_{e+f}^{LL}+C_{1}\, {\cal A}_{g+h}^{LL} \Big]
  +V_{ub}^{\ast}\,V_{uq}\,
   \Big[ a_{2}\, {\cal A}_{i+j}^{LL}+C_{1}\, {\cal A}_{k+l}^{LL} \Big]
   \nonumber \\ & & \quad
  -V_{tb}^{\ast}\,V_{tq}\,
   \Big[ (a_{3}+a_{9})\, {\cal A}_{e+f+i+j}^{LL}
        +(a_{5}+a_{7})\, {\cal A}_{e+f+i+j}^{LR}
   \nonumber \\ & & \qquad \ \ \quad
      +\ (C_{4}+C_{10})\, {\cal A}_{g+h+k+l}^{LL}
        +(C_{6}+C_{8} )\, {\cal A}_{g+h+k+l}^{LR} \Big] \Big\}
   \label{eq:amp-bd-dzdz},
   \end{eqnarray}
   \begin{eqnarray} & &
  {\cal A}(B_{q}^{{\ast}0}{\to}D_{q}^{-}D_{q}^{+})
   \nonumber \\ &=& {\cal F}\, \Big\{
   V_{cb}^{\ast}\,V_{cq}\,
   \Big[ a_{1}\, {\cal A}_{a+b}^{LL}+C_{2}\, {\cal A}_{c+d}^{LL}
       + a_{2}\, {\cal A}_{e+f}^{LL}+C_{1}\, {\cal A}_{g+h}^{LL} \Big]
   \nonumber \\ & & \quad
  -V_{tb}^{\ast}\,V_{tq}\,
   \Big[ (a_{4}+a_{10})\, {\cal A}_{a+b}^{LL}
       + (a_{6}+a_{8})\,  {\cal A}_{a+b}^{SP}
       + (a_{3}+a_{9})\,  {\cal A}_{e+f}^{LL}
   \nonumber \\ & & \quad
      +\ (C_{3}+C_{9})\, {\cal A}_{c+d}^{LL}
      +  (C_{5}+C_{7})\, {\cal A}_{c+d}^{SP}
      + (C_{4}+C_{10})\, {\cal A}_{g+h}^{LL}
   \nonumber \\ & & \quad
      +\ (a_{5}-\frac{1}{2}\,a_{7})\, {\cal A}_{i+j}^{LR}
      +  (C_{5}-\frac{1}{2}\,C_{7})\, {\cal A}_{k+l}^{SP}
      +  (C_{6}-\frac{1}{2}\,C_{8})\, {\cal A}_{k+l}^{LR}
   \nonumber \\ & & \quad
      +\ (a_{5}+a_{7})\, {\cal A}_{e+f}^{LR}
      +  (a_{3}+a_{4}-\frac{1}{2}\,a_{9}-\frac{1}{2}\,a_{10})\, {\cal A}_{i+j}^{LL}
   \nonumber \\ & & \quad
      +\ (C_{6}+C_{8})\,  {\cal A}_{g+h}^{LR}
      +  (C_{3}+C_{4}-\frac{1}{2}\,C_{9}-\frac{1}{2}\,C_{10})\, {\cal A}_{k+l}^{LL}
        \Big] \Big\}
   \label{eq:amp-bq-dqm-dqp},
   \end{eqnarray}
   \begin{eqnarray} & &
  {\cal A}(B_{d}^{{\ast}0}{\to}D^{-}D_{s}^{+})
   \nonumber \\ &=& {\cal F}\, \Big\{
   V_{cb}^{\ast}\,V_{cs}\,
   \Big[ a_{1}\, {\cal A}_{a+b}^{LL}+C_{2}\, {\cal A}_{c+d}^{LL} \Big]
  -V_{tb}^{\ast}\,V_{ts}\, \Big[ (a_{4}+a_{10})\, {\cal A}_{a+b}^{LL}
   \nonumber \\ & & \quad
      +\ (a_{4}-\frac{1}{2}\,a_{10})\, {\cal A}_{i+j}^{LL}
      +  (C_{3}-\frac{1}{2}\,C_{9})\, {\cal A}_{k+l}^{LL}
      +  (C_{5}-\frac{1}{2}\,C_{7})\, {\cal A}_{k+l}^{SP}
   \nonumber \\ & & \quad
      +\ (a_{6}+a_{8} )\, {\cal A}_{a+b}^{SP}
      +  (C_{3}+C_{9} )\, {\cal A}_{c+d}^{LL}
      +  (C_{5}+C_{7} )\, {\cal A}_{c+d}^{SP}
        \Big] \Big\}
   \label{eq:amp-bd-dm-dsp},
   \end{eqnarray}
   \begin{eqnarray} & &
  {\cal A}(B_{s}^{{\ast}0}{\to}D_{s}^{-}D^{+})
   \nonumber \\ &=& {\cal F}\, \Big\{
   V_{cb}^{\ast}\,V_{cd}\,
   \Big[ a_{1}\, {\cal A}_{a+b}^{LL}+C_{2}\, {\cal A}_{c+d}^{LL} \Big]
  -V_{tb}^{\ast}\,V_{td}\, \Big[ (a_{4}+a_{10})\, {\cal A}_{a+b}^{LL}
   \nonumber \\ & & \quad
      +\ (a_{4}-\frac{1}{2}\,a_{10})\, {\cal A}_{i+j}^{LL}
      +  (C_{3}-\frac{1}{2}\,C_{9})\,  {\cal A}_{k+l}^{LL}
      +  (C_{5}-\frac{1}{2}\,C_{7})\,  {\cal A}_{k+l}^{SP}
   \nonumber \\ & & \quad
      +\ (a_{6}+a_{8} )\, {\cal A}_{a+b}^{SP}
      +  (C_{3}+C_{9} )\, {\cal A}_{c+d}^{LL}
      +  (C_{5}+C_{7} )\, {\cal A}_{c+d}^{SP}
        \Big] \Big\}
   \label{eq:amp-bs-dsm-dp},
   \end{eqnarray}
   \begin{eqnarray} & &
  {\cal A}(B_{d}^{{\ast}0}{\to}D_{s}^{-}D_{s}^{+})
   \nonumber \\ &=& {\cal F}\, \Big\{
   V_{cb}^{\ast}\,V_{cd}\,
   \Big[ a_{2}\, {\cal A}_{e+f}^{LL} + C_{1}\, {\cal A}_{g+h}^{LL} \Big]
   -V_{tb}^{\ast}\,V_{td}\, \Big[ (a_{3}+a_{9})\, {\cal A}_{e+f}^{LL}
   \nonumber \\ & & \quad
      +\ (a_{5}+a_{7} )\, {\cal A}_{e+f}^{LR}
      +  (C_{4}+C_{10})\, {\cal A}_{g+h}^{LL}
      +  (C_{6}+C_{8} )\, {\cal A}_{g+h}^{LR}
  \nonumber \\ & & \quad
      +\ (a_{3}-\frac{1}{2}\,a_{9} )\, {\cal A}_{i+j}^{LL}
      +  (C_{4}-\frac{1}{2}\,C_{10})\, {\cal A}_{k+l}^{LL}
  \nonumber \\ & & \quad
      +\ (a_{5}-\frac{1}{2}\,a_{7} )\, {\cal A}_{i+j}^{LR}
      +  (C_{6}-\frac{1}{2}\,C_{8} )\, {\cal A}_{k+l}^{LR} \Big] \Big\}
   \label{eq:amp-bd-dsds},
   \end{eqnarray}
   \begin{eqnarray} & &
  {\cal A}(B_{s}^{{\ast}0}{\to}D^{-}D^{+})
   \nonumber \\ &=& {\cal F}\, \Big\{
   V_{cb}^{\ast}\,V_{cs}\,
   \Big[ a_{2}\, {\cal A}_{e+f}^{LL} + C_{1}\, {\cal A}_{g+h}^{LL} \Big]
   -V_{tb}^{\ast}\,V_{ts}\, \Big[ (a_{3}+a_{9})\, {\cal A}_{e+f}^{LL}
   \nonumber \\ & & \quad
      +\ (a_{5}+a_{7} )\, {\cal A}_{e+f}^{LR}
      +  (C_{4}+C_{10})\, {\cal A}_{g+h}^{LL}
      +  (C_{6}+C_{8} )\, {\cal A}_{g+h}^{LR}
  \nonumber \\ & & \quad
      +\ (a_{3}-\frac{1}{2}\,a_{9} )\, {\cal A}_{i+j}^{LL}
      +  (C_{4}-\frac{1}{2}\,C_{10})\, {\cal A}_{k+l}^{LL}
  \nonumber \\ & & \quad
      +\ (a_{5}-\frac{1}{2}\,a_{7} )\, {\cal A}_{i+j}^{LR}
      +  (C_{6}-\frac{1}{2}\,C_{8} )\, {\cal A}_{k+l}^{LR} \Big] \Big\}
   \label{eq:amp-bs-dmdp},
   \end{eqnarray}
   \begin{equation}
  {\cal F}\, =\, \sqrt{2}\,G_{F}\,\frac{{\pi}\,C_{F}}{N_{c}}\,
  f_{B_{q}^{\ast}}\, f_{\bar{D}}\, f_{D}\, m_{1}\,
  ({\epsilon}_{B_{q}^{\ast}}{\cdot}p_{\bar{D}})
   \label{eq:amp-coe},
   \end{equation}
  where $C_{i}$ is the Wilson coefficient; the parameter $a_{i}$ is defined as
  \begin{equation}
  a_{i} = \left\{ \begin{array}{l}
  C_{i}+C_{i+1}/N_{c}  \quad \text{for odd $i$}; \\
  C_{i}+C_{i-1}/N_{c}  \quad \text{for even $i$},
  \end{array} \right.
  \label{eq:ai}
  \end{equation}
  and ${\cal A}_{m_{1}+m_{2}+{\cdots}}^{n}$ is
  an abbreviation for ${\cal A}_{m_{1}}^{n}$ $+$
  ${\cal A}_{m_{2}}^{n}$ $+$ ${\cdots}$, where the
  subscript $m_{i}$ corresponds to one of indices of
  Fig.\ref{feynman}; the superscript $n$ refers
  to three possible Dirac structures ${\Gamma}_{1}{\otimes}{\Gamma}_{2}$
  of the operators $(\bar{q}_{1}q_{2})_{{\Gamma}_{1}}(\bar{q}_{3}q_{4})_{{\Gamma}_{2}}$,
  namely $n$ $=$ $LL$ for $(V-A){\otimes}(V-A)$, $n$ $=$ $LR$
  for $(V-A){\otimes}(V+A)$, and $n$ $=$ $SP$ for $-2(S-P){\otimes}(S+P)$.
  The explicit expressions of the building blocks ${\cal A}_{m_{i}}^{n}$
  are collected in Appendix.

  \section{Numerical results and discussion}
  \label{sec03}
  In the rest frame of the $B_{q}^{\ast}$ meson, the branching ratio is
  defined as
   \begin{equation}
  {\cal B}r\ =\ \frac{1}{24{\pi}}\,
   \frac{p}{m_{B^{\ast}}^{2}{\Gamma}_{B_{q}^{\ast}}}\,
  {\vert}{\cal A}(B^{\ast}{\to}\overline{D}D){\vert}^{2}
   \label{br},
   \end{equation}
  where ${\Gamma}_{B_{q}^{\ast}}$ is the full decay width of
   the $B_{q}^{\ast}$ meson.

   \begin{table}[ht]
   \caption{The numerical values of the input parameters.}
   \label{tab:input}
   \begin{ruledtabular}
   \begin{tabular}{lll}
   \multicolumn{3}{c}{CKM parameter\footnotemark[1] \cite{pdg}} \\ \hline
   \multicolumn{3}{c}{
    ${\lambda}$ $=$ $0.22506{\pm}0.00050$, \qquad
    $A$ $=$ $0.811{\pm}0.026$, \qquad
    $\bar{\rho}$ $=$ $0.124^{+0.019}_{-0.018}$, \qquad
    $\bar{\eta}$ $=$ $0.356{\pm}0.011$; } \\ \hline
    \multicolumn{3}{c}{mass and decay constant} \\ \hline
    $m_{B_{s}^{\ast}}$ $=$ $5415.4^{+1.8}_{-1.5}$ MeV \cite{pdg},
  & $f_{B_{s}^{\ast}}$ $=$ $213{\pm}7$ MeV \cite{prd91.114509},
  & ${\Lambda}^{(5)}_{\rm QCD}$ $=$ $210{\pm}14$ MeV \cite{pdg}, \\
    $m_{B_{u,d}^{\ast}}$ $=$ $5324.65{\pm}0.25$ MeV \cite{pdg},
  & $f_{B_{u,d}^{\ast}}$ $=$ $175{\pm}6$ MeV \cite{prd91.114509},
  & ${\Lambda}^{(4)}_{\rm QCD}$ $=$ $292{\pm}16$ MeV \cite{pdg}, \\
    $m_{D_{s}}$ $=$ $1968.27{\pm}0.10$ MeV \cite{pdg},
  & $f_{D_{s}}$ $=$ $249.0{\pm}1.2$ MeV \cite{pdg},
  & $m_{b}$ $=$ $4.78{\pm}0.06$ GeV \cite{pdg}, \\
    $m_{D_{d}}$ $=$ $1869.58{\pm}0.09$ MeV \cite{pdg},
  & $f_{D_{u,d}}$ $=$ $211.9{\pm}1.1$ MeV \cite{pdg},
  & $m_{c}$ $=$ $1.67{\pm}0.07$ GeV \cite{pdg}, \\
    $m_{D_{u}}$ $=$ $1864.83{\pm}0.05$ MeV \cite{pdg},
  & $m_{u,d}$ ${\simeq}$ $0.31$ GeV \cite{uds},
  & $m_{s}$ ${\simeq}$ $0.51$ GeV \cite{uds}.
   \end{tabular}
   \end{ruledtabular}
   \footnotetext[1]{The relations between the CKM parameters (${\rho}$, ${\eta}$)
   and ($\bar{\rho}$, $\bar{\eta}$) are \cite{pdg}: $({\rho}+i{\eta})$ $=$
   $\displaystyle \frac{ \sqrt{1-A^{2}{\lambda}^{4}}(\bar{\rho}+i\bar{\eta}) }
  { \sqrt{1-{\lambda}^{2}}[1-A^{2}{\lambda}^{4}(\bar{\rho}+i\bar{\eta})] }$.}
  \end{table}

  The numerical values of some input parameters are listed in
  Table \ref{tab:input}, where if it is not specified explicitly,
  their central values will be fixed as the default inputs.
  Besides, the full decay width of the $B_{q}^{\ast}$ meson,
  ${\Gamma}_{B_{q}^{\ast}}$, is also an essential parameter.
  Unfortunately, an experimental measurement on ${\Gamma}_{B_{q}^{\ast}}$
  is unavailable now, because the soft photon from the
  $B_{q}^{\ast}$ ${\to}$ $B_{q}{\gamma}$ process is usually beyond
  the detection capability of electromagnetic calorimeters
  sitting at existing high energy colliders.
  It is well known that the electromagnetic radiation process
  $B_{q}^{\ast}$ ${\to}$ $B_{q}{\gamma}$ dominates the decay of
  the $B_{q}^{\ast}$ meson.
  So, for the time being, the full decay width will be approximated
  by the radiative partial width, i.e., ${\Gamma}_{B_{q}^{\ast}}$
  ${\simeq}$ ${\Gamma}(B_{q}^{\ast}{\to}B_{q}{\gamma})$.
  At present, the information on ${\Gamma}(B_{q}^{\ast}{\to}B_{q}{\gamma})$
  comes mainly from theoretical estimation.
  Theoretically, the partial decay width of the M1 transition (spin-flip)
  process has the expression \cite{jhep1404.177,epja52.90}
   \begin{equation}
  {\Gamma}(B_{q}^{\ast}{\to}B_{q}{\gamma})\ =\
   \frac{4}{3}\,{\alpha}\, k_{\gamma}^{3}\, {\mu}^{2}_{h}
   \label{m1-width},
   \end{equation}
  where ${\alpha}$ is the fine structure constant;
  $k_{\gamma}$ $=$ $(m_{B_{q}^{\ast}}^{2}-m_{B_{q}}^{2})/2m_{B_{q}^{\ast}}$
  is the photon momentum in the rest frame of the $B_{q}^{\ast}$ meson;
  ${\mu}_{h}$ is the M1 moment of the $B_{q}^{\ast}$ meson.
  There are plenty of theoretical predictions on
  ${\Gamma}(B_{q}^{\ast}{\to}B_{q}{\gamma})$, for example,
  the numbers in Table 7 in Ref.\cite{jhep1404.177} and
  Tables 3 and 4 in  Ref.\cite{epja52.90}, but these
  estimation suffer from large uncertainties due to our
  insufficient understanding on the M1 moments of mesons.
  In principle, the M1 moment of a meson should be a
  combination of the M1 moments of the constituent quark and antiquark.
  For a heavy-light meson, the M1 moment of a heavy quark might be
  negligible relative to the M1 moment of a light quark, because
  it is widely assumed that the mass of a heavy quark is usually
  much larger than the mass of a light quark, and that the M1 moment
  is inversely proportional to the mass of a charged particle.
  With the M1 moment relations among light $u$, $d$, $s$ quarks,
  ${\vert}{\mu}_{u}{\vert}$ $>$ ${\vert}{\mu}_{d}{\vert}$ $>$
  ${\vert}{\mu}_{s}{\vert}$ \cite{uds}, one could expect to have
  ${\Gamma}(B_{u}^{\ast}{\to}B_{u}{\gamma})$ $>$
  ${\Gamma}(B_{d}^{\ast}{\to}B_{d}{\gamma})$ $>$
  ${\Gamma}(B_{s}^{\ast}{\to}B_{s}{\gamma})$,
  and so ${\Gamma}_{B_{u}^{\ast}}$ $>$ ${\Gamma}_{B_{d}^{\ast}}$
  $>$ ${\Gamma}_{B_{s}^{\ast}}$.
  Of course, more details about the width ${\Gamma}_{B_{q}^{\ast}}$
  is beyond the scope of this paper.
  In our calculation, in order to give a quantitative estimation of
  the branching ratios for the $B^{\ast}$ ${\to}$ $\overline{D}D$ decays,
  we will fix
   \begin{equation}
  {\Gamma}_{B_{u}^{\ast}}\ {\sim}\
  {\Gamma}(B_{u}^{\ast}{\to}B_{u}{\gamma})\ {\sim}\
   450\,\text{eV}
   \label{m1-width-u},
   \end{equation}
   \begin{equation}
  {\Gamma}_{B_{d}^{\ast}}\ {\sim}\
  {\Gamma}(B_{d}^{\ast}{\to}B_{d}{\gamma})\ {\sim}\
   150\,\text{eV}
   \label{m1-width-d},
   \end{equation}
   \begin{equation}
  {\Gamma}_{B_{s}^{\ast}}\ {\sim}\
  {\Gamma}(B_{s}^{\ast}{\to}B_{s}{\gamma})\ {\sim}\
   100\,\text{eV}
   \label{m1-width-s},
   \end{equation}
  which is basically consistent with the recent results
  in Refs.\cite{jhep1404.177,epja52.90}.

  In order to investigate the effects from different DA models,
  we explore three scenarios,
  \begin{itemize}
  \item Scenario I:
  ${\phi}_{B^{\ast}}^{v}$ $=$ Eq.(\ref{da-bqv}),
  ${\phi}_{B^{\ast}}^{t}$ $=$ Eq.(\ref{da-bqt}),
  ${\phi}_{\bar{D},D}^{a}$ $=$ Eq.(\ref{da-cqa})
  and ${\phi}_{\bar{D},D}^{p}$ $=$ Eq.(\ref{da-cqp}).
  \item Scenario II:
  ${\phi}_{B^{\ast}}^{v}$ $=$ ${\phi}_{B^{\ast}}^{t}$ $=$ Eq.(\ref{da-bqv}),
  and ${\phi}_{\bar{D},D}^{a}$ $=$
  ${\phi}_{\bar{D},D}^{p}$ $=$ Eq.(\ref{da-cqa}).
  \item Scenario III:
  ${\phi}_{B^{\ast}}^{v}$ $=$ ${\phi}_{B^{\ast}}^{t}$ $=$ Eq.(\ref{da-bqv}),
  ${\phi}_{\bar{D},D}^{a}$ $=$
  ${\phi}_{\bar{D},D}^{p}$ $=$ Eq.(\ref{wave-d-xb}).
  \end{itemize}

  Our numerical results are presented in Table \ref{tab:br},
  where the uncertainties come from the typical scale
  $(1{\pm}0.1)t_{i}$, mass $m_{c}$ and $m_{b}$, and the CKM
  parameters, respectively.
  The following are some comments.

   \begin{table}[ht]
   \caption{The branching ratios for the $B^{\ast}$ ${\to}$ $\overline{D}D$ decays,
   where the theoretical uncertainties come from scale $(1{\pm}0.1)t_{i}$,
   mass $m_{c}$ and $m_{b}$, and the CKM parameters, respectively; the numbers
   in columns correspond to different DA scenarios.}
   \label{tab:br}
   \begin{ruledtabular}
   \begin{tabular}{lcccc}
    & class & I & II & III \\ \hline
     ${\cal B}r(B_{u}^{{\ast}+}{\to}\overline{D}_{u}^{0}D_{d}^{+}){\times}10^{11}$
   & B
   & $7.65^{+ 1.81+ 0.04+ 0.62}_{- 0.68- 0.78- 0.59}$
   & $2.21^{+ 0.40+ 0.17+ 0.18}_{- 0.17- 0.21- 0.17}$
   & $1.24^{+ 0.20+ 0.15+ 0.10}_{- 0.09- 0.16- 0.09}$ \\
     ${\cal B}r(B_{u}^{{\ast}+}{\to}\overline{D}_{u}^{0}D_{s}^{+}){\times}10^{9}$
   & A
   & $1.89^{+ 0.45+ 0.01+ 0.14}_{- 0.17- 0.19- 0.13}$
   & $0.57^{+ 0.11+ 0.04+ 0.04}_{- 0.05- 0.05- 0.04}$
   & $0.31^{+ 0.05+ 0.04+ 0.02}_{- 0.02- 0.04- 0.02}$ \\
     ${\cal B}r(B_{d}^{{\ast}0}{\to}D_{d}^{-}D_{s}^{+}){\times}10^{9}$
   & A
   & $5.68^{+ 1.36+ 0.01+ 0.42}_{- 0.51- 0.54- 0.40}$
   & $1.72^{+ 0.33+ 0.13+ 0.13}_{- 0.14- 0.17- 0.12}$
   & $0.94^{+ 0.16+ 0.12+ 0.07}_{- 0.07- 0.12- 0.07}$ \\
     ${\cal B}r(B_{s}^{{\ast}0}{\to}D_{s}^{-}D_{d}^{+}){\times}10^{10}$
   & B
   & $6.34^{+ 1.40+ 0.07+ 0.51}_{- 0.54- 0.45- 0.49}$
   & $2.09^{+ 0.38+ 0.12+ 0.17}_{- 0.16- 0.18- 0.16}$
   & $1.15^{+ 0.19+ 0.12+ 0.09}_{- 0.08- 0.13- 0.09}$\\
     ${\cal B}r(B_{d}^{{\ast}0}{\to}D_{d}^{-}D_{d}^{+}){\times}10^{10}$
   & B
   & $2.27^{+ 0.55+ 0.00+ 0.18}_{- 0.21- 0.21- 0.17}$
   & $0.68^{+ 0.12+ 0.05+ 0.05}_{- 0.05- 0.06- 0.05}$
   & $0.39^{+ 0.06+ 0.05+ 0.03}_{- 0.03- 0.05- 0.03}$\\
     ${\cal B}r(B_{s}^{{\ast}0}{\to}D_{s}^{-}D_{s}^{+}){\times}10^{8}$
   & A
   & $1.51^{+ 0.34+ 0.03+ 0.11}_{- 0.13- 0.12- 0.11}$
   & $0.53^{+ 0.10+ 0.03+ 0.04}_{- 0.04- 0.04- 0.04}$
   & $0.30^{+ 0.05+ 0.03+ 0.02}_{- 0.02- 0.03- 0.02}$\\ \hline
     ${\cal B}r(B_{d}^{{\ast}0}{\to}\overline{D}_{u}^{0}D_{u}^{0}){\times}10^{14}$
   & D
   & $1.53^{+ 0.73+ 1.45+ 0.19}_{- 0.61- 0.21- 0.17}$
   & $0.43^{+ 0.14+ 0.53+ 0.07}_{- 0.12- 0.34- 0.06}$
   & $4.10^{+ 0.12+ 0.26+ 0.48}_{- 0.02- 0.25- 0.44}$\\
     ${\cal B}r(B_{d}^{{\ast}0}{\to}D_{s}^{-}D_{s}^{+}){\times}10^{13}$
   & D
   & $1.11^{+ 0.16+ 0.22+ 0.09}_{- 0.19- 0.20- 0.09}$
   & $0.99^{+ 0.06+ 0.17+ 0.08}_{- 0.06- 0.18- 0.08}$
   & $0.66^{+ 0.03+ 0.04+ 0.05}_{- 0.02- 0.04- 0.05}$\\
     ${\cal B}r(B_{s}^{{\ast}0}{\to}\overline{D}_{u}^{0}D_{u}^{0}){\times}10^{13}$
   & C
   & $ 7.57^{+ 3.07+ 6.57+ 0.58}_{- 2.52- 1.66- 0.55}$
   & $ 1.97^{+ 0.67+ 2.71+ 0.15}_{- 0.58- 1.68- 0.15}$
   & $10.35^{+ 0.64+ 1.00+ 0.79}_{- 0.39- 1.01- 0.76}$\\
     ${\cal B}r(B_{s}^{{\ast}0}{\to}D_{d}^{-}D_{d}^{+}){\times}10^{13}$
   & C
   & $ 7.62^{+ 3.07+ 6.51+ 0.57}_{- 2.53- 1.61- 0.55}$
   & $ 1.99^{+ 0.67+ 2.64+ 0.15}_{- 0.58- 1.67- 0.14}$
   & $10.41^{+ 0.60+ 0.98+ 0.77}_{- 0.37- 1.00- 0.74}$
   \end{tabular}
   \end{ruledtabular}
   \end{table}

  (1)
  Generally, the $B^{\ast}$ ${\to}$ $\overline{D}D$ decay modes may
  be divided into four classes. The tree contributions of classes
  A, B, C, D are proportional to the factors of
  $V_{cb}^{\ast}\,V_{cs}\,a_{1}$ ${\sim}$ $A{\lambda}^{2}\,a_{1}$,
  $V_{cb}^{\ast}\,V_{cd}\,a_{1}$ ${\sim}$ $A{\lambda}^{3}\,a_{1}$,
  $V_{cb}^{\ast}\,V_{cs}\,a_{2}$ ${\sim}$ $A{\lambda}^{2}\,a_{2}$,
  $V_{cb}^{\ast}\,V_{cd}\,a_{2}$ ${\sim}$ $A{\lambda}^{3}\,a_{2}$,
  respectively. Classes C and D are pure annihilation processes.
  There is a clear hierarchy of branching ratios, i.e.,
  ${\cal B}r(\text{class A})$ ${\gtrsim}$ $10^{-9}$,
  ${\cal B}r(\text{class B})$ ${\gtrsim}$ $10^{-11}$,
  ${\cal B}r(\text{class C})$ ${\gtrsim}$ $10^{-13}$,
  ${\cal B}r(\text{class D})$ ${\lesssim}$ $10^{-13}$.

  In addition, for each class, the magnitude relations between the decay
  constants $f_{B_{s}^{{\ast}}}$ $>$ $f_{B_{u,d}^{{\ast}}}$ and
  $f_{D_{s}}$ $>$ $f_{D_{u,d}}$, and relations among the decay widths
  ${\Gamma}_{B_{s}^{\ast}}$ $<$ ${\Gamma}_{B_{d}^{\ast}}$
  $<$ ${\Gamma}_{B_{u}^{\ast}}$, result in the size relations
  among the branching ratios, i.e.,
   \begin{equation}
  {\cal B}r(B_{s}^{{\ast}0}{\to}D_{s}^{-}D_{s}^{+})  >
  {\cal B}r(B_{d}^{{\ast}0}{\to}D_{d}^{-}D_{s}^{+})  >
  {\cal B}r(B_{u}^{{\ast}+}{\to}\overline{D}_{u}^{0}D_{s}^{+})
   \label{r-01-01},
   \end{equation}
   \begin{equation}
  {\cal B}r(B_{s}^{{\ast}0}{\to}D_{s}^{-}D_{d}^{+})  >
  {\cal B}r(B_{d}^{{\ast}0}{\to}D_{d}^{-}D_{d}^{+})  >
  {\cal B}r(B_{u}^{{\ast}+}{\to}\overline{D}_{u}^{0}D_{d}^{+})
   \label{r-01-02},
   \end{equation}
   \begin{equation}
  {\cal B}r(B_{d}^{{\ast}0}{\to}D_{s}^{-}D_{s}^{+})  >
  {\cal B}r(B_{d}^{{\ast}0}{\to}\overline{D}_{u}^{0}D_{u}^{0})
   \label{r-01-03}.
   \end{equation}

  (2)
  Due to the isospin symmetry, there are some approximate relations
  among the branching ratios, for example,
   \begin{equation}
   \frac{ {\cal B}r(B_{d}^{{\ast}0}{\to}D_{d}^{-}D_{s}^{+}) }
        { {\cal B}r(B_{u}^{{\ast}+}{\to}\overline{D}_{u}^{0}D_{s}^{+}) }
   \ {\approx}\ \frac{ {\Gamma}_{B_{u}^{\ast}} }{ {\Gamma}_{B_{d}^{\ast}} }
   \  {\approx}\ 3
   \label{r-02-01},
   \end{equation}
   \begin{equation}
   \frac{ {\cal B}r(B_{d}^{{\ast}0}{\to}D_{d}^{-}D_{d}^{+}) }
        { {\cal B}r(B_{u}^{{\ast}+}{\to}\overline{D}_{u}^{0}D_{d}^{+}) }
   \ {\approx}\ \frac{ {\Gamma}_{B_{u}^{\ast}} }{ {\Gamma}_{B_{d}^{\ast}} }
   \  {\approx}\ 3
   \label{r-02-02},
   \end{equation}
   \begin{equation}
   \frac{ {\cal B}r(B_{s}^{{\ast}0}{\to}\overline{D}_{u}^{0}D_{u}^{0}) }
        { {\cal B}r(B_{s}^{{\ast}0}{\to}D_{d}^{-}D_{d}^{+}) }
   \ {\approx}\ 1
   \label{r-02-03}.
   \end{equation}
  In addition, there are some other approximate relations,
  for example,
   \begin{equation}
   \frac{ {\cal B}r(B_{u}^{{\ast}+}{\to}\overline{D}_{u}^{0}D_{s}^{+}) }
        { {\cal B}r(B_{u}^{{\ast}+}{\to}\overline{D}_{u}^{0}D_{d}^{+}) }
   \ {\approx}\ \frac{ f_{D_{s}}^{2} }{ f_{D_{d}}^{2} }\,
   \frac{ {\vert} V_{cs} {\vert}^{2} }{ {\vert} V_{cd} {\vert}^{2} }
   \label{r-03-01},
   \end{equation}
   \begin{equation}
   \frac{ {\cal B}r(B_{d}^{{\ast}0}{\to}D_{d}^{-}D_{s}^{+}) }
        { {\cal B}r(B_{d}^{{\ast}0}{\to}D_{d}^{-}D_{d}^{+}) }
   \ {\approx}\ \frac{ f_{D_{s}}^{2} }{ f_{D_{d}}^{2} }\,
   \frac{ {\vert} V_{cs} {\vert}^{2} }{ {\vert} V_{cd} {\vert}^{2} }
   \label{r-03-02},
   \end{equation}
   \begin{equation}
   \frac{ {\cal B}r(B_{s}^{{\ast}0}{\to}D_{s}^{-}D_{s}^{+}) }
        { {\cal B}r(B_{s}^{{\ast}0}{\to}D_{s}^{-}D_{d}^{+}) }
   \ {\approx}\ \frac{ f_{D_{s}}^{2} }{ f_{D_{d}}^{2} }\,
   \frac{ {\vert} V_{cs} {\vert}^{2} }{ {\vert} V_{cd} {\vert}^{2} }
   \label{r-03-03}.
   \end{equation}

  (3)
  Our study shows that
  (i) both the emission topologies and the annihilation topologies
  contribute to the decay channels of Classes A and B. Furthermore,
  the contributions
  from the emission topologies are dominant over those from the
  annihilation topologies.
  (ii) For the pure annihilation decay channels of Classes C and D,
  the factorizable contributions are color-suppressed, so the
  nonfactorizable contributions are the main ones.
  In addition, the interferences among different topologies are
  important.
  (iii) Compared with the contributions of the tree operators, the
  contributions of the penguin operators are small because of the
  suppression from the small Wilson coefficients.
  The contributions of the topologies in Fig.\ref{feynman}(i-l) are
  much less than those of the topologies in Fig.\ref{feynman}(e-h).
  (iv) For the decay channels with the final
  states $\overline{D}_{q}D_{q}$, the contribution of the factorizable
  annihilation topology in Fig.\ref{feynman}(e) [Fig.\ref{feynman}(i)]
  is the same magnitude as that in Fig.\ref{feynman}(f) [Fig.\ref{feynman}(j)]
  due to the flavor symmetry.
  (v)
  The interferences between factorizable topologies in
  Fig.\ref{feynman}(e) and Fig.\ref{feynman}(f)
  [Fig.\ref{feynman}(i) and Fig.\ref{feynman}(j)] are constructive,
  while the interferences between nonfactorizable topologies in
  Fig.\ref{feynman}(g) and Fig.\ref{feynman}(h)
  [Fig.\ref{feynman}(k) and Fig.\ref{feynman}(l)] are destructive.

   \begin{table}[ht]
   \caption{The fractions (in the unit of \%) of the different $b$-hadron species.}
   \label{tab:bb-fr}
   \begin{ruledtabular}
   \begin{tabular}{c|ccc|ccccc}
     channels
   & $(b\bar{s})(\bar{b}s)$
   & $B_{s}^{{\ast}0}\overline{B}_{s}^{{\ast}0}$
   & $B_{s}^{{\ast}0}\overline{B}_{s}^{0}+{\rm c.c.}$
   & $~~(b\bar{u})(\bar{b}u) \atop +(b\bar{d})(\bar{b}d)$
   & $B^{\ast}\overline{B}^{\ast}$
   & $B^{\ast}\overline{B}+{\rm c.c.}$
   & $B^{\ast}\overline{B}{\pi}+{\rm c.c.}$
   & $B^{\ast}\overline{B}^{\ast}{\pi}$
     \\ \hline
     fractions
   & ${\cal B}_{s}$
   & ${\cal B}_{B_{s}^{{\ast}0}\overline{B}_{s}^{{\ast}0}}$
   & ${\cal B}_{B_{s}^{{\ast}0}\overline{B}_{s}^{0}}$
   & ${\cal B}_{u}$ $+$ ${\cal B}_{d}$
   & ${\cal B}_{B^{\ast}\overline{B}^{\ast}}$
   \\ \hline
     ${\Upsilon}(5S)$ \cite{epjc74.3026}
   & $19.5^{+3.0}_{-2.3}$
   & $17.6{\pm}2.8$
   & $1.4{\pm}0.7$
   & $73.7{\pm}6.0$
   & $37.5{\pm}3.7$
   & $13.7{\pm}1.7$
   & $7.3{\pm}2.4$
   & $1.0{\pm}1.5$
   \\
     Tevatron \cite{1412.7515}
   & $10.0{\pm}1.0$ & &
   & $70.0{\pm}4.0$ & & & &
   \end{tabular}
   \end{ruledtabular}
   \end{table}

  (4)
  The $B_{s}^{{\ast}0}$ ${\to}$ $D_{s}^{-}D_{s}^{+}$,
  $B_{d}^{{\ast}0}$ ${\to}$ $D_{d}^{-}D_{s}^{+}$,
  $B_{u}^{{\ast}+}$ ${\to}$ $\overline{D}_{u}^{0}D_{s}^{+}$
  decays, belonging to class A, have relatively large branching ratios,
  ${\cal B}r(\text{class A})$ ${\gtrsim}$ $10^{-9}$.
  The numbers of the $B_{q}^{\ast}$ mesons in a data sample can be estimated by
   \begin{equation}
   N(B_{q}^{\ast})\ =\ {\cal L}_{\rm int}\,
  {\times}\, {\sigma}_{b\bar{b}}\,{\times}\,{\cal B}_{q}\, {\times}\,
   \frac{ {\cal B}_{B_{q}^{\ast}} } { {\cal B}_{q} }
   \label{r-04-01},
   \end{equation}
   \begin{equation}
  {\cal B}_{B_{q}^{\ast}}\ =\ 2\,{\times}\,{\cal B}_{B_{q}^{\ast}\overline{B}_{q}^{\ast}}
   +2\,{\times}\,{\cal B}_{B_{q}^{\ast}\overline{B}_{q}^{\ast}{\pi}}
   +{\cal B}_{B_{q}^{\ast}\overline{B}_{q}+{\rm c.c.}}
   +{\cal B}_{B_{q}^{\ast}\overline{B}_{q}{\pi}+{\rm c.c}}
   +{\cdots}
   \label{r-04-02},
   \end{equation}
  where ${\cal L}_{\rm int}$ is the integrated luminosity,
  ${\sigma}_{b\bar{b}}$ denotes the $b\bar{b}$ pair production cross section,
  ${\cal B}_{q}$ refers to the fragmentation fraction of $(b\bar{q})(\bar{b}q)$ events,
  and ${\cal B}_{B_{q}^{\ast}\overline{B}_{q}^{\ast}}$,
  ${\cal B}_{B_{q}^{\ast}\overline{B}_{q}^{\ast}X}$, ....
  represent the production fractions of specific modes (see Table \ref{tab:bb-fr}).
  With a large production cross section of the process $e^{+}e^{-}$ ${\to}$
  $b\bar{b}$ at the ${\Upsilon}(5S)$ peak ${\sigma}_{b\bar{b}}$ $=$
  $(0.340{\pm}0.016)\,{\rm nb}$ \cite{epjc74.3026}
  and a high luminosity $8{\times}10^{35}\,{\rm cm}^{-2}\,{\rm s}^{-1}$
  at the forthcoming SuperKEKB \cite{1002.5012}, it is expected that
  some $3.3{\times}10^{9}$ $B_{u,d}^{\ast}$ and $1.2{\times}10^{9}$
  $B_{s}^{\ast}$ mesons could be available per $10\,{\rm ab}^{-1}$
  ${\Upsilon}(5S)$ dataset, corresponding to a few events of the
  $B_{d}^{{\ast}0}$ ${\to}$ $D_{d}^{-}D_{s}^{+}$ and $B_{u}^{{\ast}+}$
  ${\to}$ $\overline{D}_{u}^{0}D_{s}^{+}$ decays and dozens of the
  $B_{s}^{{\ast}0}$ ${\to}$ $D_{s}^{-}D_{s}^{+}$ decay.
  At high energy hadron colliders, for example, with a visible
  $b$-hadron cross section at the LHCb ${\sigma}_{b\bar{b}}$
  ${\sim}$ $100\,{\rm {\mu}b}$ \cite{pdg,crp16.435},
  a similar ratio ${\cal B}_{q}$ at Tevatron and a similar ratio
  ${\cal B}_{B_{q}^{\ast}}/{\cal B}_{q}$ at ${\Upsilon}(5S)$,
  some $9.2{\times}10^{13}$ $B_{u,d}^{\ast}$
  and $1.9{\times}10^{13}$ $B_{s}^{\ast}$ events per ${\rm ab}^{-1}$
  dataset should be available at the LHCb, corresponding to more than
  $10^{5}$ (class A) $B_{s}^{{\ast}0}$ ${\to}$ $D_{s}^{-}D_{s}^{+}$,
  $B_{d}^{{\ast}0}$ ${\to}$ $D_{d}^{-}D_{s}^{+}$, and
  $B_{u}^{{\ast}+}$ ${\to}$ $\overline{D}_{u}^{0}D_{s}^{+}$
  decay events and over $10^{4}$ (class B)
  $B_{s}^{{\ast}0}$ ${\to}$ $D_{s}^{-}D_{d}^{+}$,
  $B_{d}^{{\ast}0}$ ${\to}$ $D_{d}^{-}D_{d}^{+}$
  $B_{u}^{{\ast}+}$ ${\to}$ $\overline{D}_{u}^{0}D_{d}^{+}$ decay events,
  which are measurable at the future LHCb experiments.
  However, the search for pure annihilation processes (classes
  C and D) at LHC and SuperKEKB should still be very challenging.

  (5)
  Compared with the branching ratios ${\gtrsim}$ $10^{-5}$ for the $B$
  ${\to}$ $\overline{D}D$ decays \cite{pdg,prd81.034006}, the branching
  ratios for the $B^{\ast}$ ${\to}$ $\overline{D}D$ decays are smaller by
  at least three orders of magnitude. This fact might imply that
  the background from the $B^{\ast}$ ${\to}$ $\overline{D}D$ decays
  could be safely neglected when one analyzes the $B$ ${\to}$
  $\overline{D}D$ decays, but not vice versa, i.e., one of main
  pollution backgrounds for the $B^{\ast}$ ${\to}$ $\overline{D}D$
  decays would come from the $B$ ${\to}$ $\overline{D}D$ decays,
  even if the invariant mass of the $\overline{D}D$ meson pair
  could be used to distinguish the $B^{\ast}$ meson from the $B$ meson
  experimentally.

  (6)
  For the $B_{q}^{\ast}$ decays of classes A and B, our estimation of
  the branching ratios agrees well with that based on the naive
  factorization approach \cite{1605.01630}.
  One of the important reasons is that these processes are all
  $a_{1}$-dominated (color favored), and in general, insensitive
  to nonfactorizable corrections to the HTME.
  Of course, one fact is clear that there are many theoretical
  uncertainties, especially, regarding the discrepancy among different DA
  scenarios, which results from our uncertain knowledge of
  the long-distance QCD effects and the underlying dynamics
  of low energy hadron interactions.
  Moreover, as aforementioned, there are large uncertainties
  of the decay width ${\Gamma}_{B_{q}^{\ast}}$. With a different value
  of ${\Gamma}_{B_{q}^{\ast}}$, the branching ratios
  in Table \ref{tab:br} should be multiplied by the factors of
  ${450\,{\rm eV}}/{{\Gamma}_{B_{u}^{\ast}}}$,
  ${150\,{\rm eV}}/{{\Gamma}_{B_{d}^{\ast}}}$,
  ${100\,{\rm eV}}/{{\Gamma}_{B_{s}^{\ast}}}$
  for the $B_{u}^{\ast}$, $B_{d}^{\ast}$, $B_{s}^{\ast}$
  decays, respectively.
  In addition, many other factors, such as the final state
  interactions, models for the $B^{\ast}$ and $D$ meson wave functions\footnotemark[2],
  \footnotetext[2]{In principle, one can do a global fit on the $B^{\ast}$
  and $D$ meson wave functions with experimental measurements in the future,
  analogous to that with the ${\chi}^{2}$ method in Ref.\cite{prd78.014018}.
  The fitting will be a very time-consuming work,
  because the amplitudes for the $B^{\ast}$ ${\to}$ $\overline{D}D$ decays
  are expressed as the multidimensional integral with the pQCD approach.
  In addition, there is no measurement report on the $B^{\ast}$ ${\to}$
  $\overline{D}D$ decays at the moment.}
  higher order corrections to the HTME, and so on, are not
  carefully scrutinized here, but deserve much dedicated
  study. Our estimation may be just an order of magnitude.

  \section{Summary}
  \label{sec04}
  With the running LHC and the forthcoming SuperKEKB, a large amount of
  $B^{\ast}$ data should be in stock soon, which will make it seemingly
  possible to explore the $B^{\ast}$ weak decays experimentally.
  A theoretical study is necessary in order to offer
  a timely reference, and is helpful in clearing up some of
  puzzles surrounding heavy meson weak decays.
  In this paper, we investigated the $B^{\ast}$ ${\to}$
  $\overline{D}D$ decays with the phenomenological pQCD approach.
  It is found that the $B_{s}^{{\ast}0}$ ${\to}$ $D_{s}^{-}D_{s}^{+}$,
  $B_{d}^{{\ast}0}$ ${\to}$ $D_{d}^{-}D_{s}^{+}$, and
  $B_{u}^{{\ast}+}$ ${\to}$ $\overline{D}_{u}^{0}D_{s}^{+}$
  decays have branching ratios ${\gtrsim}$ $10^{-9}$, and
  will be promisingly accessible at the future high luminosity
  experiments, with help of a sophisticated experimental analytical
  technique to effectively suppress or exclude the background
  from the $B$ ${\to}$ $\overline{D}D$ decays.

  \section*{Acknowledgments}
  The work is supported by the National Natural Science Foundation
  of China (Grant Nos. U1632109, 11547014 and 11475055).

  \begin{appendix}
  \section{Amplitude building blocks for $B^{\ast}$ ${\to}$ $\overline{D}D$ decays}
  \label{block}
   \begin{eqnarray}
  {\cal A}_{a}^{LL} &=&
  {\int}_{0}^{1}dx_{1}
  {\int}_{0}^{1}dx_{2}
  {\int}_{0}^{\infty}b_{1}db_{1}
  {\int}_{0}^{\infty}b_{2}db_{2}\,
  {\alpha}_{s}(t_{a})\,
  H_{ef}({\alpha}_{e},{\beta}_{a},b_{1},b_{2})
   \nonumber \\ && \!\!\!\! \!\!\!\! \!\!\!\!
  {\times}\, E_{ef}(t_{a})\, {\phi}_{B^{\ast}}^{v}(x_{1})\,
   \Big\{ {\phi}_{\bar{D}}^{a}(x_{2})\,
   ( m_{1}^{2}\,\bar{x}_{2}+m_{3}^{2}\,x_{2} )
   + {\phi}_{\bar{D}}^{p}(x_{2})\, m_{2}\,m_{b} \Big\}
   \label{amp:a-ll},
   \end{eqnarray}
   \begin{eqnarray}
  {\cal A}_{a}^{SP} &=&
  -2\,m_{3}\,
  {\int}_{0}^{1}dx_{1}
  {\int}_{0}^{1}dx_{2}
  {\int}_{0}^{\infty}b_{1}db_{1}
  {\int}_{0}^{\infty}b_{2}db_{2}\,
  H_{ef}({\alpha}_{e},{\beta}_{a},b_{1},b_{2})
   \nonumber \\ &{\times}&
  {\alpha}_{s}(t_{a})\, E_{ef}(t_{a})\,
  {\phi}_{B^{\ast}}^{v}(x_{1})\,
   \Big\{ {\phi}_{\bar{D}}^{a}(x_{2})\,m_{b}
   + {\phi}_{\bar{D}}^{p}(x_{2})\, m_{2}\,x_{2} \Big\}
   \label{amp:a-sp},
   \end{eqnarray}
   \begin{eqnarray}
  {\cal A}_{b}^{LL} &=&
  {\int}_{0}^{1}dx_{1}
  {\int}_{0}^{1}dx_{2}
  {\int}_{0}^{\infty}b_{1}db_{1}
  {\int}_{0}^{\infty}b_{2}db_{2}\,
  {\alpha}_{s}(t_{b})\,
  H_{ef}({\alpha}_{e},{\beta}_{a},b_{2},b_{1})
   \nonumber \\ &{\times}&
  E_{ef}(t_{b})\,\Big\{ {\phi}_{B^{\ast}}^{t}(x_{1})\, \Big[
  {\phi}_{\bar{D}}^{p}(x_{2})\, 2\,m_{1}\,m_{2}\,\bar{x}_{1}
 -{\phi}_{\bar{D}}^{a}(x_{2})\,m_{1}\,m_{c} \Big]
   \nonumber \\ &+&
  {\phi}_{B^{\ast}}^{v}(x_{1})\, \Big[
  {\phi}_{\bar{D}}^{p}(x_{2})\, 2\,m_{2}\,m_{c}
 -{\phi}_{\bar{D}}^{a}(x_{2})\,
  ( m_{2}^{2}\,\bar{x}_{1}+m_{3}^{2}\,x_{1} ) \Big] \Big\}
   \label{amp:b-ll},
   \end{eqnarray}
   \begin{eqnarray}
  {\cal A}_{b}^{SP} &=&
  2\,m_{3}\,
  {\int}_{0}^{1}dx_{1}
  {\int}_{0}^{1}dx_{2}
  {\int}_{0}^{\infty}b_{1}db_{1}
  {\int}_{0}^{\infty}b_{2}db_{2}\,
  {\alpha}_{s}(t_{b})\,
  H_{ef}({\alpha}_{e},{\beta}_{a},b_{2},b_{1})
   \nonumber \\ & &  \!\!\!\! \!\!
   \!\!\!\! \!\!\!\! \!\!\!\! \!\!\!\! {\times}\,
  E_{ef}(t_{b})\,\Big\{ {\phi}_{B^{\ast}}^{v}(x_{1})\, \Big[
  {\phi}_{\bar{D}}^{a}(x_{2})\, m_{c}
 -{\phi}_{\bar{D}}^{p}(x_{2})\, 2\,m_{2} \Big]
 +{\phi}_{B^{\ast}}^{t}(x_{1})\,{\phi}_{\bar{D}}^{a}(x_{2})\,m_{1}\,x_{1} \Big\}
   \label{amp:b-sp},
   \end{eqnarray}
   \begin{eqnarray}
  {\cal A}_{c}^{LL} &=&
   \frac{1}{N_{c}}\,
  {\int}_{0}^{1}dx_{1}
  {\int}_{0}^{1}dx_{2}
  {\int}_{0}^{1}dx_{3}
  {\int}_{0}^{\infty}db_{1}
  {\int}_{0}^{\infty}b_{2}db_{2}
  {\int}_{0}^{\infty}b_{3}db_{3}\,
  {\alpha}_{s}(t_{c})\,
  E_{n}(t_{c})
   \nonumber \\ &{\times}&
  H_{en}({\alpha}_{e},{\beta}_{c},b_{3},b_{2},b_{1})\,
  \Big\{ {\phi}_{B^{\ast}}^{t}(x_{1})\,
  {\phi}_{\bar{D}}^{p}(x_{2})\,
  {\phi}_{D}^{a}(x_{3})\, m_{1}\, m_{2}\,(x_{1}-x_{2})
   \nonumber \\ & & +
  {\phi}_{B^{\ast}}^{v}(x_{1})\,
  {\phi}_{\bar{D}}^{a}(x_{2})\, \Big[
  {\phi}_{D}^{a}(x_{3})\,s\,(x_{2}-\bar{x}_{3})
 -{\phi}_{D}^{p}(x_{3})\, m_{3}\, m_{c} \Big] \Big\}
   \label{amp:c-ll},
   \end{eqnarray}
   \begin{eqnarray}
  {\cal A}_{c}^{SP} &=&
   \frac{1}{N_{c}}\,
  {\int}_{0}^{1}dx_{1}
  {\int}_{0}^{1}dx_{2}
  {\int}_{0}^{1}dx_{3}
  {\int}_{0}^{\infty}db_{1}
  {\int}_{0}^{\infty}b_{2}db_{2}
  {\int}_{0}^{\infty}b_{3}db_{3}\,
  H_{en}({\alpha}_{e},{\beta}_{c},b_{3},b_{2},b_{1})
   \nonumber \\ &{\times}&
   E_{n}(t_{c})\, {\alpha}_{s}(t_{c})\,
  \Big\{ {\phi}_{B^{\ast}}^{v}(x_{1})\,
  {\phi}_{\bar{D}}^{p}(x_{2})\, m_{2}\, \Big[
  {\phi}_{D}^{p}(x_{3})\,  m_{3}\,(x_{2}-\bar{x}_{3})
 -{\phi}_{D}^{a}(x_{3})\,  m_{c} \Big]
   \nonumber \\ & & +
  {\phi}_{B^{\ast}}^{t}(x_{1})\,
  {\phi}_{\bar{D}}^{a}(x_{2})\, m_{1}\, \Big[
  {\phi}_{D}^{a}(x_{3})\, m_{c}
 +{\phi}_{D}^{p}(x_{3})\, m_{3}\, (\bar{x}_{3}-x_{1})
   \Big] \Big\}
   \label{amp:c-sp},
   \end{eqnarray}
   \begin{eqnarray}
  {\cal A}_{d}^{LL} &=&
   \frac{1}{N_{c}}\,
  {\int}_{0}^{1}dx_{1}
  {\int}_{0}^{1}dx_{2}
  {\int}_{0}^{1}dx_{3}
  {\int}_{0}^{\infty}db_{1}
  {\int}_{0}^{\infty}b_{2}db_{2}
  {\int}_{0}^{\infty}b_{3}db_{3}\,
  {\alpha}_{s}(t_{d})\,
  E_{n}(t_{d})
   \nonumber \\ &{\times}&
  H_{en}({\alpha}_{e},{\beta}_{d},b_{3},b_{2},b_{1})\,
  {\phi}_{D}^{a}(x_{3})\,
   \Big\{ {\phi}_{B^{\ast}}^{t}(x_{1})\,
  {\phi}_{\bar{D}}^{p}(x_{2})\, m_{1}\,m_{2}\,(x_{1}-x_{2})
   \nonumber \\ & & +
   {\phi}_{B^{\ast}}^{v}(x_{1})\,
  {\phi}_{\bar{D}}^{a}(x_{2})\,
   (2\,m_{2}^{2}\,x_{2}+s\,x_{3}-t\,x_{1}) \Big\}
   \label{amp:d-ll},
   \end{eqnarray}
   \begin{eqnarray}
  {\cal A}_{d}^{SP} &=&
   \frac{1}{N_{c}}\,
  {\int}_{0}^{1}dx_{1}
  {\int}_{0}^{1}dx_{2}
  {\int}_{0}^{1}dx_{3}
  {\int}_{0}^{\infty}db_{1}
  {\int}_{0}^{\infty}b_{2}db_{2}
  {\int}_{0}^{\infty}b_{3}db_{3}\,
  {\alpha}_{s}(t_{d})\,
  E_{n}(t_{d})
   \nonumber \\ &{\times}&
  H_{en}({\alpha}_{e},{\beta}_{d},b_{3},b_{2},b_{1})\,
  {\phi}_{D}^{p}(x_{3})\,
   \Big\{ {\phi}_{B^{\ast}}^{v}(x_{1})\,
  {\phi}_{\bar{D}}^{p}(x_{2})\, m_{2}\,m_{3}\,(x_{3}-x_{2})
   \nonumber \\ & & +
   {\phi}_{B^{\ast}}^{t}(x_{1})\,
  {\phi}_{\bar{D}}^{a}(x_{2})\,
   m_{1}\,m_{3}\,(x_{1}-x_{3}) \Big\}
   \label{amp:d-sp},
   \end{eqnarray}
   \begin{eqnarray}
  {\cal A}_{e}^{LL} &=& {\cal A}_{e}^{LR}\ =\
  {\int}_{0}^{1}dx_{2}
  {\int}_{0}^{1}dx_{3}
  {\int}_{0}^{\infty}b_{2}db_{2}
  {\int}_{0}^{\infty}b_{3}db_{3}\,
  H_{af}({\alpha}_{q},{\beta}_{e},b_{2},b_{3})
   \nonumber \\ &{\times}&
  {\alpha}_{s}(t_{e})\, E_{af}(t_{e})\, \Big\{
  {\phi}_{D}^{p}(x_{3})\,2\,m_{3}\, \Big[
  {\phi}_{\bar{D}}^{a}(x_{2})\,m_{c}
 +{\phi}_{\bar{D}}^{p}(x_{2})\,m_{2}\,\bar{x}_{2} \Big]
   \nonumber \\ & &
 -{\phi}_{D}^{a}(x_{3})\, \Big[
  {\phi}_{\bar{D}}^{a}(x_{2})\,(m_{1}^{2}\,x_{2}+m_{3}^{2}\bar{x}_{2} )
 +{\phi}_{\bar{D}}^{p}(x_{2})\,m_{2}\,m_{c} \Big] \Big\}
   \label{amp:e-ll},
   \end{eqnarray}
   \begin{eqnarray}
  {\cal A}_{f}^{LL} &=& {\cal A}_{f}^{LR}\ =\
  {\int}_{0}^{1}dx_{2}
  {\int}_{0}^{1}dx_{3}
  {\int}_{0}^{\infty}b_{2}db_{2}
  {\int}_{0}^{\infty}b_{3}db_{3}\,
  H_{af}({\alpha}_{q},{\beta}_{f},b_{3},b_{2})
   \nonumber \\ &{\times}&
  {\alpha}_{s}(t_{f})\, E_{af}(t_{f})\, \Big\{
  {\phi}_{\bar{D}}^{p}(x_{2})\,2\,m_{2}\, \Big[
  {\phi}_{D}^{a}(x_{3})\,m_{c}
 +{\phi}_{D}^{p}(x_{3})\,m_{3}\,\bar{x}_{3} \Big]
   \nonumber \\ & &
 -{\phi}_{\bar{D}}^{a}(x_{2})\, \Big[
  {\phi}_{D}^{a}(x_{3})\,(m_{1}^{2}\,x_{3}+m_{2}^{2}\bar{x}_{3} )
 +{\phi}_{D}^{p}(x_{3})\,m_{3}\,m_{c} \Big] \Big\}
   \label{amp:f-ll},
   \end{eqnarray}
   \begin{eqnarray}
  {\cal A}_{g}^{LL} &=&
   \frac{1}{N_{c}}\,
  {\int}_{0}^{1}dx_{1}
  {\int}_{0}^{1}dx_{2}
  {\int}_{0}^{1}dx_{3}
  {\int}_{0}^{\infty}b_{1}db_{1}
  {\int}_{0}^{\infty}b_{2}db_{2}
  {\int}_{0}^{\infty}db_{3}\,
  H_{an}({\alpha}_{q},{\beta}_{g},b_{1},b_{2},b_{3})
   \nonumber \\ &{\times}&
   E_{n}(t_{g})\, {\alpha}_{s}(t_{g})\,
   \Big\{ {\phi}_{\bar{D}}^{a}(x_{2})\, {\phi}_{D}^{a}(x_{3})\,
   \Big[ {\phi}_{B^{\ast}}^{v}(x_{1})\,
  (s\,x_{2}+2\,m_{3}^{2}\,x_{3}-u\,\bar{x}_{1})
   \nonumber \\ & &
  +{\phi}_{B^{\ast}}^{t}(x_{1})\,m_{1}\,m_{b} \Big]
  +{\phi}_{B^{\ast}}^{v}(x_{1})\, {\phi}_{\bar{D}}^{p}(x_{2})\,
  {\phi}_{D}^{p}(x_{3})\, m_{2}\,m_{3}\,(x_{2}-x_{3}) \Big\}
   \label{amp:g-LL},
   \end{eqnarray}
   \begin{eqnarray}
  {\cal A}_{g}^{LR} &=&
   \frac{1}{N_{c}}\,
  {\int}_{0}^{1}dx_{1}
  {\int}_{0}^{1}dx_{2}
  {\int}_{0}^{1}dx_{3}
  {\int}_{0}^{\infty}b_{1}db_{1}
  {\int}_{0}^{\infty}b_{2}db_{2}
  {\int}_{0}^{\infty}db_{3}\,
  H_{an}({\alpha}_{q},{\beta}_{g},b_{1},b_{2},b_{3})
   \nonumber \\ &{\times}&
   E_{n}(t_{g})\, {\alpha}_{s}(t_{g})\,
   \Big\{ {\phi}_{\bar{D}}^{a}(x_{2})\, {\phi}_{D}^{a}(x_{3})\,
   \Big[ {\phi}_{B^{\ast}}^{v}(x_{1})\,
  (t\,\bar{x}_{1}-2\,m_{2}^{2}\,x_{2}-s\,x_{3})
   \nonumber \\ & &
  -{\phi}_{B^{\ast}}^{t}(x_{1})\,m_{1}\,m_{b} \Big]
  +{\phi}_{B^{\ast}}^{v}(x_{1})\, {\phi}_{\bar{D}}^{p}(x_{2})\,
  {\phi}_{D}^{p}(x_{3})\, m_{2}\,m_{3}\,(x_{2}-x_{3}) \Big\}
   \label{amp:g-LR},
   \end{eqnarray}
   \begin{eqnarray}
  {\cal A}_{h}^{LL} &=&
   \frac{1}{N_{c}}\,
  {\int}_{0}^{1}dx_{1}
  {\int}_{0}^{1}dx_{2}
  {\int}_{0}^{1}dx_{3}
  {\int}_{0}^{\infty}b_{1}db_{1}
  {\int}_{0}^{\infty}b_{2}db_{2}
  {\int}_{0}^{\infty}db_{3}\,
  E_{n}(t_{h})
   \nonumber \\ &{\times}&
   {\alpha}_{s}(t_{h})\,{\phi}_{B^{\ast}}^{v}(x_{1})\, \Big\{
  {\phi}_{\bar{D}}^{a}(x_{2})\, {\phi}_{D}^{a}(x_{3})\,
  (2\,m_{2}^{2}\,x_{2}+s\,x_{3}-t\,x_{1})
   \nonumber \\ &+&
  {\phi}_{\bar{D}}^{p}(x_{2})\, {\phi}_{D}^{p}(x_{3})\,
  m_{2}\,m_{3}\,(x_{3}-x_{2}) \Big\}\,
  H_{an}({\alpha}_{q},{\beta}_{h},b_{1},b_{2},b_{3})
   \label{amp:h-LL},
   \end{eqnarray}
   \begin{eqnarray}
  {\cal A}_{h}^{LR} &=&
   \frac{1}{N_{c}}\,
  {\int}_{0}^{1}dx_{1}
  {\int}_{0}^{1}dx_{2}
  {\int}_{0}^{1}dx_{3}
  {\int}_{0}^{\infty}b_{1}db_{1}
  {\int}_{0}^{\infty}b_{2}db_{2}
  {\int}_{0}^{\infty}db_{3}\,
  E_{n}(t_{h})
   \nonumber \\ &{\times}&
   {\alpha}_{s}(t_{h})\,{\phi}_{B^{\ast}}^{v}(x_{1})\, \Big\{
  {\phi}_{\bar{D}}^{a}(x_{2})\, {\phi}_{D}^{a}(x_{3})\,
  (u\,x_{1}-s\,x_{2}- 2\,m_{3}^{2}\,x_{3})
   \nonumber \\ &+&
  {\phi}_{\bar{D}}^{p}(x_{2})\, {\phi}_{D}^{p}(x_{3})\,
  m_{2}\,m_{3}\,(x_{3}-x_{2}) \Big\}\,
  H_{an}({\alpha}_{q},{\beta}_{h},b_{1},b_{2},b_{3})
   \label{amp:h-LR},
   \end{eqnarray}
   \begin{eqnarray}
  {\cal A}_{i}^{LL} &=& {\cal A}_{i}^{LR}\ =\
  {\int}_{0}^{1}dx_{2}
  {\int}_{0}^{1}dx_{3}
  {\int}_{0}^{\infty}b_{2}db_{2}
  {\int}_{0}^{\infty}b_{3}db_{3}\,
  {\alpha}_{s}(t_{i})\,
  H_{af}({\alpha}_{c},{\beta}_{i},b_{2},b_{3})
   \nonumber \\ && \!\!\!\! \!\!\!\! \!\!\!\!
   \!\!\!\! \!\!\!\! {\times}\, E_{af}(t_{i})\, \Big\{
  {\phi}_{\bar{D}}^{a}(x_{2})\, {\phi}_{D}^{a}(x_{3})\,
  (m_{1}^{2}\,\bar{x}_{2}+m_{3}^{2}\,x_{2})
 -{\phi}_{\bar{D}}^{p}(x_{2})\,{\phi}_{D}^{p}(x_{3})\,
  2\,m_{2}\,m_{3}\,x_{2} \Big\}
   \label{amp:i-ll},
   \end{eqnarray}
   \begin{eqnarray}
  {\cal A}_{j}^{LL} &=& {\cal A}_{j}^{LR}\ =\
  {\int}_{0}^{1}dx_{2}
  {\int}_{0}^{1}dx_{3}
  {\int}_{0}^{\infty}b_{2}db_{2}
  {\int}_{0}^{\infty}b_{3}db_{3}\,
  {\alpha}_{s}(t_{j})\,
  H_{af}({\alpha}_{c},{\beta}_{j},b_{3},b_{2})
   \nonumber \\ && \!\!\!\! \!\!\!\! \!\!\!\!
   \!\!\!\! \!\!\!\! {\times}\, E_{af}(t_{j})\, \Big\{
  {\phi}_{\bar{D}}^{a}(x_{2})\, {\phi}_{D}^{a}(x_{3})\,
  (m_{1}^{2}\,\bar{x}_{3}+m_{2}^{2}\,x_{3})
 -{\phi}_{\bar{D}}^{p}(x_{2})\,{\phi}_{D}^{p}(x_{3})\,
  2\, m_{2}\,m_{3}\,x_{3} \Big\}
   \label{amp:j-ll},
   \end{eqnarray}
   \begin{eqnarray}
  {\cal A}_{k}^{LL} &=&
   \frac{1}{N_{c}}\,
  {\int}_{0}^{1}dx_{1}
  {\int}_{0}^{1}dx_{2}
  {\int}_{0}^{1}dx_{3}
  {\int}_{0}^{\infty}b_{1}db_{1}
  {\int}_{0}^{\infty}b_{2}db_{2}
  {\int}_{0}^{\infty}db_{3}\,
  H_{an}({\alpha}_{c},{\beta}_{k},b_{1},b_{2},b_{3})
   \nonumber \\ &{\times}&
   E_{n}(t_{k})\, {\alpha}_{s}(t_{k})\,
   \Big\{ {\phi}_{\bar{D}}^{a}(x_{2})\, {\phi}_{D}^{a}(x_{3})\,
   \Big[ {\phi}_{B^{\ast}}^{v}(x_{1})\,
  (2\,m_{2}^{2}\,x_{2}+s\,x_{3}-t\,x_{1})
   \nonumber \\ & &
  -{\phi}_{B^{\ast}}^{t}(x_{1})\,m_{1}\,m_{b} \Big]
  +{\phi}_{B^{\ast}}^{v}(x_{1})\, {\phi}_{\bar{D}}^{p}(x_{2})\,
  {\phi}_{D}^{p}(x_{3})\, m_{2}\,m_{3}\,(x_{3}-x_{2}) \Big\}
   \label{amp:k-LL},
   \end{eqnarray}
   \begin{eqnarray}
  {\cal A}_{k}^{LR} &=&
   \frac{1}{N_{c}}\,
  {\int}_{0}^{1}dx_{1}
  {\int}_{0}^{1}dx_{2}
  {\int}_{0}^{1}dx_{3}
  {\int}_{0}^{\infty}b_{1}db_{1}
  {\int}_{0}^{\infty}b_{2}db_{2}
  {\int}_{0}^{\infty}db_{3}\,
  H_{an}({\alpha}_{c},{\beta}_{k},b_{1},b_{2},b_{3})
   \nonumber \\ &{\times}&
   E_{n}(t_{k})\, {\alpha}_{s}(t_{k})\,
   \Big\{ {\phi}_{\bar{D}}^{a}(x_{2})\, {\phi}_{D}^{a}(x_{3})\,
   \Big[ {\phi}_{B^{\ast}}^{v}(x_{1})\,
  (u\,x_{1}-s\,x_{2}-2\,m_{3}^{2}\,x_{3})
   \nonumber \\ & &
  +{\phi}_{B^{\ast}}^{t}(x_{1})\,m_{1}\,m_{b} \Big]
  +{\phi}_{B^{\ast}}^{v}(x_{1})\, {\phi}_{\bar{D}}^{p}(x_{2})\,
  {\phi}_{D}^{p}(x_{3})\, m_{2}\,m_{3}\,(x_{3}-x_{2}) \Big\}
   \label{amp:k-LR},
   \end{eqnarray}
   \begin{eqnarray}
  {\cal A}_{k}^{SP} &=&
   \frac{1}{N_{c}}\,
  {\int}_{0}^{1}dx_{1}
  {\int}_{0}^{1}dx_{2}
  {\int}_{0}^{1}dx_{3}
  {\int}_{0}^{\infty}b_{1}db_{1}
  {\int}_{0}^{\infty}b_{2}db_{2}
  {\int}_{0}^{\infty}db_{3}\,
  H_{an}({\alpha}_{c},{\beta}_{k},b_{1},b_{2},b_{3})
   \nonumber \\ &{\times}&
   E_{n}(t_{k})\,
   \Big\{ {\phi}_{\bar{D}}^{a}(x_{2})\, {\phi}_{D}^{p}(x_{3})\,
   m_{3}\, \Big[ {\phi}_{B^{\ast}}^{v}(x_{1})\, m_{b}
   +{\phi}_{B^{\ast}}^{t}(x_{1})\,m_{1}\,(x_{1}-x_{3}) \Big]
   \nonumber \\ & &
  +{\phi}_{\bar{D}}^{p}(x_{2})\, {\phi}_{D}^{a}(x_{3})\,
   m_{2}\, \Big[ {\phi}_{B^{\ast}}^{v}(x_{1})\,m_{b}
  +{\phi}_{B^{\ast}}^{t}(x_{1})\,m_{1}\,(x_{1}-x_{2})
    \Big] \Big\}\, {\alpha}_{s}(t_{k})
   \label{amp:k-SP},
   \end{eqnarray}
   \begin{eqnarray}
  {\cal A}_{l}^{LL} &=&
   \frac{1}{N_{c}}\,
  {\int}_{0}^{1}dx_{1}
  {\int}_{0}^{1}dx_{2}
  {\int}_{0}^{1}dx_{3}
  {\int}_{0}^{\infty}b_{1}db_{1}
  {\int}_{0}^{\infty}b_{2}db_{2}
  {\int}_{0}^{\infty}db_{3}\,
  E_{n}(t_{l})
   \nonumber \\ &{\times}&
  {\alpha}_{s}(t_{l})\, {\phi}_{B^{\ast}}^{v}(x_{1})\,
   \Big\{ {\phi}_{\bar{D}}^{a}(x_{2})\, {\phi}_{D}^{a}(x_{3})\,
  (u\,x_{1}-s\,\bar{x}_{2}-2\,m_{3}^{2}\,\bar{x}_{3})
   \nonumber \\ &+&
  {\phi}_{\bar{D}}^{p}(x_{2})\, {\phi}_{D}^{p}(x_{3})\,
  m_{2}\,m_{3}\,(x_{2}-x_{3}) \Big\}\,
  H_{an}({\alpha}_{c},{\beta}_{l},b_{1},b_{2},b_{3})
   \label{amp:l-LL},
   \end{eqnarray}
   \begin{eqnarray}
  {\cal A}_{l}^{LR} &=&
   \frac{1}{N_{c}}\,
  {\int}_{0}^{1}dx_{1}
  {\int}_{0}^{1}dx_{2}
  {\int}_{0}^{1}dx_{3}
  {\int}_{0}^{\infty}b_{1}db_{1}
  {\int}_{0}^{\infty}b_{2}db_{2}
  {\int}_{0}^{\infty}db_{3}\,
  E_{n}(t_{l})
   \nonumber \\ &{\times}&
  {\alpha}_{s}(t_{l})\, {\phi}_{B^{\ast}}^{v}(x_{1})\,
   \Big\{ {\phi}_{\bar{D}}^{a}(x_{2})\, {\phi}_{D}^{a}(x_{3})\,
  (2\,m_{2}^{2}\,\bar{x}_{2}+s\,\bar{x}_{3}-t\,x_{1})
   \nonumber \\ &+&
  {\phi}_{\bar{D}}^{p}(x_{2})\, {\phi}_{D}^{p}(x_{3})\,
  m_{2}\,m_{3}\,(x_{2}-x_{3}) \Big\}\,
  H_{an}({\alpha}_{c},{\beta}_{l},b_{1},b_{2},b_{3})
   \label{amp:l-LR},
   \end{eqnarray}
   \begin{eqnarray}
  {\cal A}_{l}^{SP} &=&
  \frac{m_{1}}{N_{c}} \,
  {\int}_{0}^{1}dx_{1}
  {\int}_{0}^{1}dx_{2}
  {\int}_{0}^{1}dx_{3}
  {\int}_{0}^{\infty}b_{1}db_{1}
  {\int}_{0}^{\infty}b_{2}db_{2}
  {\int}_{0}^{\infty}db_{3}
   \nonumber \\ &{\times}& E_{n}(t_{l})\,
  {\alpha}_{s}(t_{l})\, {\phi}_{B^{\ast}}^{t}(x_{1})\,
   \Big\{ {\phi}_{\bar{D}}^{a}(x_{2})\, {\phi}_{D}^{p}(x_{3})\,
   m_{3}\, (\bar{x}_{3}-x_{1})
   \nonumber \\ &+&
  {\phi}_{\bar{D}}^{p}(x_{2})\, {\phi}_{D}^{a}(x_{3})\,
   m_{2}\,(\bar{x}_{2}-x_{1}) \Big\}\,
  H_{an}({\alpha}_{c},{\beta}_{l},b_{1},b_{2},b_{3})
   \label{amp:l-SP},
   \end{eqnarray}
  where the subscript $i$ of ${\cal A}_{i}^{j}$ corresponds to
  the indices of Fig.\ref{feynman}; the superscript $j$ refers
  to three possible Dirac structures ${\Gamma}_{1}{\otimes}{\Gamma}_{2}$
  of the operators $(\bar{q}_{1}q_{2})_{{\Gamma}_{1}}(\bar{q}_{3}q_{4})_{{\Gamma}_{2}}$,
  namely $j$ $=$ $LL$ for $(V-A){\otimes}(V-A)$, $j$ $=$ $LR$
  for $(V-A){\otimes}(V+A)$, and $j$ $=$ $SP$ for $-2(S-P){\otimes}(S+P)$.

  The function $H_{i}$ and the Sudakov factor $E_{i}$ are defined as
   \begin{equation}
   H_{ef}({\alpha}_{e},{\beta},b_{m},b_{n})\, =\,
   K_{0}(b_{m}\sqrt{-{\alpha}_{e}})\, \Big\{
   {\theta}(b_{m}-b_{n}) K_{0}(b_{m}\sqrt{-{\beta}})\,
   I_{0}(b_{n}\sqrt{-{\beta}})
   + (b_{m} {\leftrightarrow} b_{n}) \Big\}
   \label{amp:hef},
   \end{equation}
   \begin{eqnarray}
  H_{en}({\alpha}_{e},{\beta},b_{3},b_{2},b_{1}) &=&
  \Big\{ {\theta}(-{\beta})\, K_{0}(b_{3}\sqrt{-{\beta}})
  +\frac{{\pi}}{2}\,
  {\theta}({\beta})\, \Big[ i\,J_{0}(b_{3}\sqrt{{\beta}})
   - Y_{0}(b_{3}\sqrt{{\beta}}) \Big] \Big\}
   \nonumber \\ && \!\!\!\! \!\!\!\! \!\!\!\! \!\!\!\! \!\!\!\!  {\times}\,
   \Big\{ {\theta}(b_{2}-b_{3})\, K_{0}(b_{2}\sqrt{-{\alpha}_{e}})\,
   I_{0}(b_{3}\sqrt{-{\alpha}_{e}}) + (b_{2} {\leftrightarrow} b_{3})
   \Big\}\, {\delta}(b_{1}-b_{2})
   \label{amp:hen},
   \end{eqnarray}
   \begin{eqnarray}
   H_{af}({\alpha},{\beta},b_{m},b_{n})
   &=&
   \Big\{ {\theta}(b_{m}-b_{n}) \Big[ i\,J_{0}(b_{m}\sqrt{{\beta}})
   - Y_{0}(b_{m}\sqrt{{\beta}}) \Big]
   J_{0}(b_{n}\sqrt{{\beta}})
   \nonumber \\ & &
  +\, (b_{m} {\leftrightarrow} b_{n}) \Big\}\,
   \Big\{ i\,J_{0}(b_{n}\sqrt{{\alpha}})
   - Y_{0}(b_{n}\sqrt{{\alpha}}) \Big\}\,
   \frac{{\pi}^{2}}{4}
   \label{amp:haf},
   \end{eqnarray}
   \begin{eqnarray}
   H_{an}({\alpha},{\beta},b_{1},b_{2},b_{3})
   &=& \frac{\pi}{2} \,
    \Big\{ {\theta}(-{\beta})\, K_{0}(b_{1}\sqrt{-{\beta}})
  + \frac{\pi}{2}{\theta}({\beta})\,
    \Big[ i\,J_{0}(b_{1}\sqrt{{\beta}})
   - Y_{0}(b_{1}\sqrt{{\beta}}) \Big] \Big\}
   \nonumber \\ && \!\!\!\! \!\!\!\! \!\!\!\! \!\!\!\! \!\!\!\!
   \!\!\!\! \!\!\!\! \!\!\!\! \!\!\!\! \!\!\!\! {\times}\,
   \Big\{ {\theta}(b_{1}-b_{2}) \Big[ i\,J_{0}(b_{1}\sqrt{{\alpha}})
   - Y_{0}(b_{1}\sqrt{{\alpha}}) \Big]
   J_{0}(b_{2}\sqrt{{\alpha}})
   + (b_{1} {\leftrightarrow} b_{2}) \Big\}\,
  {\delta}(b_{2}-b_{3})
   \label{amp:han},
   \end{eqnarray}
   \begin{equation}
   E_{ef}(t)\ =\ {\exp}\{ -S_{B^{\ast}}(t)-S_{\bar{D}}(t) \}
   \label{sudakov-ef},
   \end{equation}
   \begin{equation}
   E_{af}(t)\ =\ {\exp}\{ -S_{\bar{D}}(t)-S_{D}(t) \}
   \label{sudakov-af},
   \end{equation}
   \begin{equation}
   E_{n}(t)\ =\ {\exp}\{ -S_{B^{\ast}}(t)-S_{\bar{D}}(t)-S_{D}(t) \}
   \label{sudakov-n},
   \end{equation}
   \begin{equation}
  S_{B^{\ast}}(t)\, =\, s(x_{1},b_{1},p_{1}^{+})
  +2{\int}_{1/b_{1}}^{t}\frac{d{\mu}}{\mu}{\gamma}_{q}
   \label{sudakov-bq},
   \end{equation}
  \begin{equation}
  S_{\bar{D}}(t)\, =\, s(x_{2},b_{2},p_{2}^{+}) + s(\bar{x}_{2},b_{2},p_{2}^{+})
  +2{\int}_{1/b_{2}}^{t}\frac{d{\mu}}{\mu}{\gamma}_{q}
   \label{sudakov-cm},
   \end{equation}
  \begin{equation}
  S_{D}(t)\, =\, s(x_{3},b_{3},p_{3}^{+}) + s(\bar{x}_{3},b_{3},p_{3}^{+})
  +2{\int}_{1/b_{3}}^{t}\frac{d{\mu}}{\mu}{\gamma}_{q}
   \label{sudakov-cp},
   \end{equation}
  where the subscript $i$ $=$ $ef$, $en$, $af$, $an$ corresponds
  to the factorizable emission topologies, the nonfactorizable emission
  topologies, the factorizable annihilation topologies, and the nonfactorizable
  annihilation topologies, respectively;
  $I_{0}$, $J_{0}$, $K_{0}$ and $Y_{0}$ are Bessel functions;
  ${\gamma}_{q}$ $=$ $-{\alpha}_{s}/{\pi}$ is the quark anomalous
  dimension; the expression of $s(x,b,Q)$ can be found in of
  Ref.\cite{pqcd1};
  ${\alpha}$ and ${\beta}$ are the virtualities of gluon and quarks.
  the subscript of ${\beta}_{i}$ corresponds to the indices of
  Fig.\ref{feynman}.
  The definitions of the particle virtuality and typical
  scale $t_{i}$ are given as follows.
   \begin{equation}
  {\alpha}_{e}\ =\ x_{1}^{2}\,m_{1}^{2}+x_{2}^{2}\,m_{2}^{2}-x_{1}\,x_{2}\,t
   \label{amp:ae},
   \end{equation}
   \begin{equation}
  {\alpha}_{q}\ =\ x_{2}^{2}\, m_{2}^{2}+x_{3}^{2}\,m_{3}^{2}+x_{2}\,x_{3}\,s
   \label{amp:aq},
   \end{equation}
   \begin{equation}
  {\alpha}_{c}\ =\ \bar{x}_{2}^{2}\,m_{2}^{2}+\bar{x}_{3}^{2}\,m_{3}^{2}
  +\bar{x}_{2}\,\bar{x}_{3}\,s
   \label{amp:ac},
   \end{equation}
   \begin{equation}
  {\beta}_{a}\ =\ m_{1}^{2} + x_{2}^{2}\,m_{2}^{2}-x_{2}\,t-m_{b}^{2}
   \label{amp:ba},
   \end{equation}
   \begin{equation}
  {\beta}_{b}\ =\ m_{2}^{2}+x_{1}^{2}\,m_{1}^{2}-x_{1}\,t - m_{c}^{2}
   \label{amp:bb},
   \end{equation}
   \begin{equation}
  {\beta}_{c}\ =\ {\alpha}_{e} + \bar{x}_{3}^{2}\,m_{3}^{2}
  -x_{1}\,\bar{x}_{3}\,u +x_{2}\,\bar{x}_{3}\,s -m_{c}^{2}
   \label{amp:bc},
   \end{equation}
   \begin{equation}
  {\beta}_{d}\ =\ {\alpha}_{e}+x_{3}^{2}\,m_{3}^{2}-x_{1}\,x_{3}\,u+x_{2}\,x_{3}\,s
   \label{amp:bd},
   \end{equation}
   \begin{equation}
  {\beta}_{e}\ =\ m_{3}^{2}+x_{2}^{2}\,m_{2}^{2}+x_{2}\,s-m_{c}^{2}\ >\ 0
   \label{amp:be},
   \end{equation}
   \begin{equation}
  {\beta}_{f}\ =\ m_{2}^{2}+x_{3}^{2}\,m_{3}^{2}+x_{3}\,s-m_{c}^{2}\ >\ 0
   \label{amp:bf},
   \end{equation}
   \begin{equation}
  {\beta}_{g}\ =\ {\alpha}_{q}+\bar{x}_{1}^{2}\,m_{1}^{2}
               - \bar{x}_{1}\,x_{2}\,t - \bar{x}_{1}\,x_{3}\,u -m_{b}^{2}
   \label{amp:bg},
   \end{equation}
   \begin{equation}
  {\beta}_{h}\ =\ {\alpha}_{q}+x_{1}^{2}\,m_{1}^{2}
               - x_{1}\,x_{2}\,t - x_{1}\,x_{3}\,u
   \label{amp:bh},
   \end{equation}
   \begin{equation}
  {\beta}_{i}\ =\ m_{3}^{2}+\bar{x}_{2}^{2}\,m_{2}^{2}+\bar{x}_{2}\,s\ >\ 0
   \label{amp:bi},
   \end{equation}
   \begin{equation}
  {\beta}_{j}\ =\ m_{2}^{2}+\bar{x}_{3}^{2}\,m_{3}^{2}+\bar{x}_{3}\,s\ >\ 0
   \label{amp:bj},
   \end{equation}
   \begin{equation}
  {\beta}_{k}\ =\ {\alpha}_{c}+\bar{x}_{1}^{2}\,m_{1}^{2}
               - \bar{x}_{1}\,\bar{x}_{2}\,t
               - \bar{x}_{1}\,\bar{x}_{3}\,u -m_{b}^{2}
   \label{amp:bk},
   \end{equation}
   \begin{equation}
  {\beta}_{l}\ =\ {\alpha}_{c}+x_{1}^{2}\,m_{1}^{2}
               - x_{1}\,\bar{x}_{2}\,t - x_{1}\,\bar{x}_{3}\,u
   \label{amp:bl},
   \end{equation}
   \begin{equation}
   t_{a,b}\ =\ {\max}\{ \sqrt{-{\alpha}_{e}},\sqrt{{\vert}{\beta}_{a,b}{\vert}},1/b_{1},1/b_{2} \}
   \label{amp:tab},
   \end{equation}
   \begin{equation}
   t_{c,d}\ =\ {\max}\{ \sqrt{-{\alpha}_{e}},\sqrt{{\vert}{\beta}_{c,d}{\vert}},1/b_{2},1/b_{3} \}
   \label{amp:tcd},
   \end{equation}
   \begin{equation}
   t_{e,f}\ =\ {\max}\{ \sqrt{{\alpha}_{q}},\sqrt{{\vert}{\beta}_{e,f}{\vert}},1/b_{2},1/b_{3} \}
   \label{amp:tef},
   \end{equation}
   \begin{equation}
   t_{g,h}\ =\ {\max}\{ \sqrt{{\alpha}_{q}},\sqrt{{\vert}{\beta}_{g,h}{\vert}},1/b_{1},1/b_{2} \}
   \label{amp:tgh},
   \end{equation}
   \begin{equation}
   t_{i,j}\ =\ {\max}\{ \sqrt{{\alpha}_{c}},\sqrt{{\vert}{\beta}_{i,j}{\vert}},1/b_{2},1/b_{3} \}
   \label{amp:tij},
   \end{equation}
   \begin{equation}
   t_{k,l}\ =\ {\max}\{ \sqrt{{\alpha}_{c}},\sqrt{{\vert}{\beta}_{k,l}{\vert}},1/b_{1},1/b_{2} \}
   \label{amp:tkl}.
   \end{equation}
  \end{appendix}

  
  \end{document}